\documentclass[runningheads]{llncs}
\usepackage[T1]{fontenc}
\usepackage{graphicx}
\usepackage{amsmath,amssymb}
\usepackage{booktabs}
\usepackage{rotating}
\usepackage{caption}
\usepackage{subcaption}
\usepackage{color}
\usepackage{float}
\begin{document}
\title{Critical density for network reconstruction}
%
%
\author{Andrea Gabrielli\inst{1,2} \and
Valentina Macchiati\inst{3,4}\and
Diego Garlaschelli\inst{4,5}}
\authorrunning{A. Gabrielli et al.}
%
\institute{Dipartimento di Ingegneria, Universit\`a degli Studi Roma Tre, Rome, Italy
\and
Enrico Fermi Research Center (CREF), Rome, Italy\and 
Scuola Normale Superiore, Pisa, Italy
\and
IMT School of Advanced Studies, Lucca, Italy\and
Lorentz Institute for Theoretical Physics, University of Leiden, The Netherlands}
\maketitle              
\begin{abstract}
The structure of many financial networks is protected by privacy and has to be inferred from aggregate observables. Here we consider one of the most successful network reconstruction methods, producing random graphs with desired link density and where the observed constraints (related to the market size of each node) are replicated as averages over the graph ensemble, but not in individual realizations. We show that there is a minimum critical link density below which the method exhibits an `unreconstructability' phase where at least one of the constraints, while still reproduced on average, is far from its expected value in typical individual realizations.
We establish the scaling of the critical density for various theoretical and empirical distributions of interbank assets and liabilities, showing that the threshold differs from the critical densities for the onset of the giant component and of the unique component in the graph.
We also find that, while dense networks are always reconstructable, sparse networks are unreconstructable if their structure is homogeneous, while they can display a crossover to reconstructability if they have an appropriate core-periphery or heterogeneous structure.
Since the reconstructability of interbank networks is related to market clearing, our results suggest that central bank interventions aimed at lowering the density of links should take network structure into account to avoid unintentional liquidity crises where the supply and demand of all financial institutions cannot be matched simultaneously.
\end{abstract}

\section{Introduction}
The interactions between the components of social, biological and economic systems are frequently unknown. 
In the case of financial networks, where nodes represent financial institutions (such as banks) and links represent credit relationships (such as loans), only aggregate exposures are observable due to confidentiality issues~\cite{bardoscia2021physics}. 
This means that only the total exposure of each node towards the aggregate of all other nodes is known. The field of \textit{network reconstruction} is interested in devising methods that make the best use of the available partial information to infer the original network \cite{squartini2018reconstruction}.
Among the probabilistic reconstruction techniques, particularly successful are those based on the maximum entropy principle~\cite{squartini2017maximum}. By maximizing the entropy, which is an information-theoretic notion of uncertainty encoded in a probability distribution~\cite{thomas2006elements}, these models generate the least biased distribution of random network configurations consistent with the constraints derived from the observed aggregate information. 
Clearly, the goodness of the reconstruction depends on the choice of the constraints: certain constraints are more informative, while others are less. 
Different implementations of the maximum entropy principle, originating from different choices of constraints, have been considered~\cite{squartini2018reconstruction,mastrandrea2014enhanced,garlaschelli2008maximum,cimini2015systemic,squartini2017network}. As a common aspect, these methods are \textit{ensemble} ones as they generate not a single, but a whole set of random configurations that are compatible on average with the available data. 

In this work we introduce the problem of \textit{reconstructability} of the network under ensemble methods. 
Reconstructability is achieved when all the constraints, besides being reproduced on average, are also `sufficiently close' to their expected value in individual typical realizations of the ensemble. 
We will first define this notion rigorously and then explore the role of the available empirical properties in making the network more or less easily reconstructable. 
Using both analytical calculations and empirical data, we study how the realized values of the constraints fluctuate around the empirical values as a function of network density. 
We find a minimum critical network density which is typically safely exceeded in the unrealistically dense regime, but not necessarily in the sparse regime of interest for real-world financial networks. 
If the network density is lower than the critical threshold, in typical configurations of the reconstructed networks the realized constraints are displaced away from the desired, observed values. 
These displacements are critical from the point of view of systemic risk. Indeed, if any of the many network-based models for financial shock propagation \cite{bardoscia2015debtrank,barucca2020network,caccioli2018network,eisenberg2001systemic,macchiati2022systemic,di2018network} (which are quite sensitive to the  underlying topology of the network) is run on realizations of a network reconstruction method, a misalignment between the realized  constraints and the empirical ones could severely bias the model-based estimation of systemic risk.
Moreover, since reconstructability can be understood as related to decentralized market clearing (where all constraints are met simultaneously via purely pairwise matchings), the lack thereof might have adverse implications in terms of illiquidity of the interbank system. 
Therefore the possibility of a crossover to unreconstructability suggests that regulatory authorities should monitor the density of interbank relationships with an increased awareness.

The rest of the paper is structured as follows: Sec.~\ref{sec:setting} introduces the general formalism and method, Sec.~\ref{sec:cases} illustrates some reference cases that are analytically tractable, Sec.~\ref{sec:loops} makes a technical check of the role played by self-loops to warrant the validity of the general approach, Sec.~\ref{sec:data} looks at empirical data from the Bankscope dataset and uses them to extrapolate the results to realistic regimes, and finally Sec.~\ref{sec:conclu} provides final remarks and conclusions.

\section{General setting\label{sec:setting}}
\subsection{The reconstruction method\label{reconstruction}}
Links in financial networks correspond to exposures, e.g. they indicate the amount of money that a bank has borrowed from another bank in a certain time window. Individual links are unobservable, while the sum of the outgoing link weights and the sum of the incoming link weights for each node are both observable, as they represent the total interbank assets and the total interbank liabilities that can be derived from the balance sheet of the corresponding institution. 
While these relationships are clearly directional, for simplicity in this paper we will frame the problem within the context of undirected networks, where we interpret each link as representing the bilateral exposure (the sum of the exposure in both directions) between two nodes. 
This does not change the essence of the phenomenology we address in this paper, i.e. the identification of a regime where the margins of the empirical network cannot be properly replicated in individual reconstructed networks, even if they are still replicated on average.

We denote the weights of the links of the (unobserved) original network as $\{w^*_{ij}\}_{ij}$ (where $w^*_{ij}=w^*_{ji}>0$ denotes the amount of the bilateral exposure between nodes $i$ and $j$, while $w^*_{ij}=0$ denotes the absence of a link) and we arrange these weights into a $N\times N$ symmetric matrix $\mathbf{W}^*$. We assume that, while we cannot observe $\mathbf{W}^*$, we can observe its margins, i.e. the so-called \emph{strength}
\begin{equation}
s^*_{i}\equiv\sum_{j=1}^Nw^*_{ij}
\end{equation}
of each node $i$ ($i=1,N$).
This quantity represents the sum of the interbank assets and the interbank liabilities for node $i$. 
Let us denote with $\vec{s}^*$ the $N$-dimensional vector of entries $\{s_i^*\}$, i.e. the (observable) \emph{strength sequence} of the (unobservable) empirical network.

Clearly, inferring $\mathbf{W}^*$ from $\vec{s}^*$ with certainty is impossible, and this is the main drawback of deterministic reconstruction methods that identify a single possible solution to the reconstruction problem~\cite{squartini2018reconstruction}. 
By contrast, probabilistic methods look for a solution, given $\vec{s}^*$, in terms of a probability distribution $P(\mathbf{A})$ over the set $\{\mathbf{A}\}$ of all $N\times N$ symmetric binary adjacency matrices (where the entry $a_{ij}$ of one such matrix is $a_{ij}=1$ if a link between nodes $i$ and $j$ is present, and $a_{ij}=0$ otherwise).
This ensures that the unobserved adjacency matrix $\mathbf{A}^*$ corresponding to the unobserved weighted matrix $\mathbf{W}^*$ (where $a^*_{ij}=1$ if $w^*_{ij}>0$ and $a^*_{ij}=0$ if $w^*_{ij}=0$) is a member of this ensemble and is therefore assigned a non-zero probability $P(\mathbf{A}^*)$. Given each generated adjacency matrix, a procedure to assign a weight to each realized link must also be designed~\cite{squartini2018reconstruction}.
This step turns the random ensemble of binary matrices into a random ensemble of weighted matrices where the entry $w_{ij}$ of a generic matrix $\mathbf{W}$ is a random variable, not to be confused with the deterministic (unknown) value $w^*_{ij}$.
Similarly, the strength $s_i=\sum_{j=1}^Nw_{ij}$ is a random variable, not to be confused with the deterministic (known) value $s^*_i$, and finally the total link weight $W=\sum_{i}\sum_{j<i}w_{ij}=\sum_{i=1}^Ns_i/2$ is a random variable, not to be confused with the total (known) weight $W^*=\sum_{i=1}^Ns^*_i/2$.

In particular, we consider a successful reconstruction method, proposed in various variants ~\cite{cimini2015systemic,cimini2015estimating,parisi2020faster}, where the probability $P(\mathbf{A})$ is factorized over pairs of nodes, i.e. the edges of the graph are assumed to be independent Bernoulli random variables. Specifically, a link between node $i$ and node $j$ is established with probability
\begin{equation}
p_{ij}(z)=\frac{z\,s^*_i\,s^*_j}{1+z\,s^*_i\,s^*_j},
\label{p}
\end{equation}
where $z$ is the only free parameter (since $\vec{s}^*$ is known) and is chosen in order to tune the resulting expected link density
\begin{equation}
d(z)=\frac{2}{N(N-1)}\sum_{i=1}^N\sum_{j<i}p_{ij}(z)
\label{density}
\end{equation}
of the reconstructed networks. 
The specific functional form of the connection probability in Eq.~\eqref{p} derives from a maximum-entropy construction where, morally, one enforces the degree (i.e. the number of links) of each node as a constraint to be met as an ensemble average~\cite{squartini2017maximum}. In such a construction, the combined quantity $x_i\equiv \sqrt{z }s^*_i$ in Eq.~\eqref{p} is in principle unrelated to the strength, as it is technically a transformed Lagrange multiplier required to enforce the degree of node $i$. However, since the degree itself is typically not observable in financial networks, the quantity $x_i$ cannot be determined from the data. The core of the `fitness ansatz' is the observation that, for a few networks whose structure has been analysed and for which the value of $x_i$ has been calculated from the empirical degrees, this value has been found to display a strong linear correlation with the empirical strength $s^*_i$ of the corresponding node~\cite{cimini2015systemic}. This linear correlation suggests that, for networks with unobservable topology, the undetermined value of $x_i$ can be assumed to be proportional (by a factor $\sqrt{z}$) to the observable strength $s^*_i$, thereby giving rise to Eq.~\eqref{p}. In this way, the only free parameter is $z$ and its effect is that of controlling for the overall link density.
Now, the empirical density is also not necessarily known, however the method allows for the exploration of a range of realistic densities as a function of $z$, based for instance on published analyses of networks of the same type for which the empirical density has been documented. Indeed, several real financial networks are found to be sparse, which means that their empirical density scales as the inverse of $N$~\cite{barucca2020network}. This implies that a choice for $z$ could be $z=z_\mathrm{sparse}$ where $z_\mathrm{sparse}$ is such that
\begin{equation}
d(z_\mathrm{sparse})\simeq \frac{k}{N},\quad k>1
\label{sparse}
\end{equation}
with $k$ finite. Clearly, the value $z_\mathrm{sparse}$ realizing the above condition depends on the entire strength sequence, and in general on the strength distribution when $N\to\infty$.

Once generated, each binary adjacency matrix $\mathbf{A}$ drawn from the ensemble is `dressed' with link weights.
In the simplest specifications of the model ~\cite{cimini2015systemic,cimini2015estimating}, the link weights are deterministic functions of the observable margins: if the link is realized, the random variable $w_{ij}$ is assigned a value $\frac{s^*_is^*_j}{2W^*p_{ij}(z)}$, otherwise it is given a zero value. This means that $w_{ij}$ is a Bernoulli random variable given by
\begin{equation}
{w}_{ij}=\left\{\begin{array}{ll}
\displaystyle{\frac{s^*_i\,s^*_j}{2W^*\,p_{ij}(z)}}&\textrm{ with probability}\quad p_{ij}(z),\\
0&\textrm{ with probability}\quad 1-p_{ij}(z).
\end{array}\right.
\label{w}
\end{equation}
A variant of this approach where the (conditional) weights on the realized links are placed not deterministically, but following a second random process resulting in exponentially distributed weights, with expected value given again by $\frac{s^*_is^*_j}{2W^*p_{ij}(z)}$, has also been considered~\cite{parisi2020faster}. While this approach is found to be superior from the point of view of the reconstruction of the unobserved weight matrix (as $\mathbf{W}^*$ is always generated with positive likelihood), our main focus here is the (simpler) reconstruction of the margins $\vec{s}^*$ of $\mathbf{W}^*$ in typical realizations, for which we can use the specification given by Eq.~\eqref{w} without loss of generality (exponentially distributed weights around the values considered here do not change the essence of our results).

Equation~\eqref{w} ensures that the (unconditional) expected weight of the link connecting nodes $i$ and $j$ equals
\begin{equation}
\langle w_{ij}\rangle=p_{ij}(z)\frac{s^*_i\,s^*_j}{2W^*\,p_{ij}(z)}=\frac{s^*_i\,s^*_j}{2W^*},
\label{expw}
\end{equation}
so that the expected strength $\langle s_i\rangle$ of each node $i$ equals the corresponding observed strength $s^*_i$ (which is a prerequisite for a successful reconstruction):
\begin{equation}
\langle s_i\rangle=\sum_{j=1}^N\langle w_{ij}\rangle=\sum_{j=1}^N\frac{s^*_i\,s^*_j}{2W^*}=s^*_i\quad\forall i.
\label{eq:sexp}
\end{equation}
Note that, in the above summations, it is crucial that $j$ takes also the value $i$ and that the `diagonal' expected value is equal to 
\begin{equation}
\langle w_{ii}\rangle= \frac{(s^*_i)^2}{2W^*},     
\label{looppaccio}
\end{equation}
i.e. it must be described by Eq.~\eqref{expw} (with $i=j$) just like any other `non-diagonal' expected weight. 
This means that, actually, the ensemble of binary adjacency matrices should allow for self-loops: the diagonal entry $a_{ii}$ should also be a Bernoulli random variable with probability $p_{ii}(z)$ given by Eq.~\eqref{p} with $i=j$, and similarly the entry $w_{ii}$ should also follow Eq.~\eqref{w} with $i=j$.
Therefore, even if the empirical (unobserved) matrices $\mathbf{A}^*$ and $\mathbf{W}^*$ have no self-loops (as it makes no sense to say that a bank lends or borrows from itself), the method needs the generation of self-loops with appropriate probabilities and weights in order to ensure that Eq.~\eqref{eq:sexp} holds and that all strengths are exactly replicated on average. 
What is important for practical purposes is that the self-loops in the reconstructed ensemble retain a negligible expected weight, so that their existence does not cause any relevant difference with respect to their absence in the real matrix $\mathbf{W}^*$. In Sec.~\ref{sec:loops} we will check that this condition is met for the cases of relevance for real-world financial networks, and for the moment assume that self-loops can be safely added to the model as they will not play a crucial role. 

In what follows, we are interested in studying various properties of the model as a function of $N$, and eventually in the thermodynamic limit $N\to\infty$. We assume that, as $N$ increases, the average empirical strength $\overline{s^*}=N^{-1}\sum_{i=1}^Ns^*_i$ remains finite.
This accounts for the fact that, irrespective of how many banks enter the market, each bank retains a finite value of assets and liabilities.
Again, this regime is consistent with the sparse regime typically observed for real-world networks.
We therefore choose units such that
\begin{equation}
\overline{s^*}=1,\qquad 2W^*=N\overline{s^*}=N,
\end{equation}
i.e. divide each empirical node strength by the average strength over all nodes.
Later, we will consider different empirical distributions of the node strengths with average value given by $\overline{s^*}=1$.
Besides its simplicity, this choice of units has the advantage that, by construction, the average value of the connection probability $p_{ij}(z)$ given by Eq.~\eqref{p} over all pairs of nodes, which is precisely the link density given by Eq.~\eqref{density}, is of order $z/(1+z)$. 
This implies that, in order to realize the sparsity condition in Eq.~\eqref{sparse}, a necessary condition is  \begin{equation}
z_\mathrm{sparse}\to 0^+\quad \mathrm{for} \quad N\to+\infty.
\label{vanish}
\end{equation}
The necessary condition, i.e. the specific speed with which $z_\mathrm{sparse}$ has to decay as $N$ grows, depends on the particular strength distribution, as we will discuss for explicit examples later.

\subsection{Transition from reconstructability to unreconstructability}
The above model ensures that the expected value of each constraint matches the corresponding observed value exactly, i.e. $\langle \vec{s}\rangle=\vec{s}^*$, however it does not ensure that all constraints can be met in each realization of the network. 
This is normal for any canonical ensemble where the constraints are by construction allowed to fluctuate around their expected values~\cite{squartini2017maximum} and is not undesirable, as long as all constraints are \emph{close enough} to their observed values in a \emph{typical} realization of the network.
This means that, for the network to be satisfactorily reconstructed, a necessary condition is that the \emph{relative fluctuation} (defined as the ratio of the standard deviation to the expected value) of the strength of each node vanishes sufficiently fast in the limit of large $N$.
In other words, we want to avoid the undesired situation where the typical realizations violate some of the constraints by an unacceptable amount, even though the expected value of each of them still coincides with the desired, observed value.

For the unobserved network $\mathbf{W}^*$ to be reconstructable from the observed strength sequence $\vec{s}^*$ we therefore require that, for each node $i$, the relative strength fluctuations
\begin{equation}
\delta_i(z)\equiv\frac{\sqrt{v_i(z)}}{\langle s_i\rangle}=\frac{\sqrt{v_i(z)}}{s^*_i}
\label{eq:relative_data}
\end{equation}
decay at least as fast as $1/\sqrt{N^\beta}$, where $v_i(z)\equiv \textrm{Var}[s_i]$ is the variance of $s_i$ and $\beta>0$ is some desired exponent. The `canonical' case is $\beta=1$, although one might in principle allow for more general scenarios.
As we now show, this requirement implies that $z$ should be larger than a critical value $z_c$, i.e. that the expected network density $d(z)$ defined in Eq.~\eqref{density} exceeds a critical threshold $d_c\equiv d(z_c)$.

Note that if $z=+\infty$ then $d(+\infty)=1$ and the model has a deterministic outcome producing only one fully connected network where all strengths are replicated exactly, with zero variance $v(+\infty)=0$.
At the opposite extreme, if $z=0$ then $d(0)=0$ and the model is again deterministic, but the output is now a single, empty network where none of the strengths are replicated as they are all zero with zero variance $v(0)=0$.
The critical value $z_c$ we are looking for is an intermediate value separating a reconstructable phase from an unreconstructable one, corresponding to a critical scaling for the largest relative fluctuation (as we want all node strengths to be satisfactorily replicated) given by
\begin{equation}
\max_{i}\{\delta_i(z_c)\}=\max_{i}\frac{\sqrt{v_i(z_c)}}{s^*_i}\equiv\sqrt{cN^{-\beta}}
\label{delta_zc}
\end{equation}
where $c>0$ is a finite constant. Since we want $\max_{i}\{\delta_i(z_c)\}< 1$ (so that the standard deviation of the strength does not exceed the expected strength) for all values of $\beta>0$ including values arbitrarily close to $0$, we also require $c<1$. 

To identify $z_c$ we first compute the variance of the weight $w_{ij}$ as
\begin{eqnarray}
\textrm{Var}[w_{ij}]&=&\langle w^2_{ij}\rangle-\langle w_{ij}\rangle^2\nonumber\\
&=&p_{ij}(z)\frac{(s^*_i\,s^*_j)^2}{(2W^*)^2\,p^2_{ij}(z)}-\langle w_{ij}\rangle^2\nonumber\\
&=&\left(\frac{s^*_i\,s^*_j}{2W^*}\right)^2\left(\frac{1}{p_{ij}(z)}-1\right)\nonumber\\
&=&\left(\frac{s^*_i\,s^*_j}{2W^*}\right)^2\frac{1}{z\, s^*_i\,s^*_j}\nonumber\\
&=&\frac{s^*_i\,s^*_j}{(2W^*)^2\,z}\nonumber\\
&=&\frac{s^*_i\,s^*_j}{N^2\,z}.
\end{eqnarray}
Next, using the independence of different edges in the graph, we compute the variance $v_i(z)$ of the strength $s_i$ as
\begin{equation}
v_i(z)=\sum_{j=1}^N\textrm{Var}[w_{ij}]=\frac{s^*_i}{2W^*z}=\frac{s^*_i}{Nz}
\end{equation}
and the resulting relative fluctuations as
\begin{equation}
\delta_i(z)\equiv\frac{\sqrt{{s^*_i}/{Nz}}}{s^*_i}=\frac{1}{\sqrt{N\,z\, s^*_i}}.
\label{relative}
\end{equation}
Clearly, the largest fluctuation is attained by the node with minimum strength:
\begin{equation}
\max_{i}\{\delta_i(z)\}=\frac{1}{\sqrt{Nz\min_{i}\{s^*_i\}}}.
\label{delta_zmax}
\end{equation}
Now, imposing that $\max_{i}\{\delta_i(z_c)\}$ equals the critical expression in Eq.~\eqref{delta_zc} implies that the critical value for $z$ is 
\begin{equation}
z_c=
\frac{N^{\beta-1}}{c\,\textrm{min}_i\{s^*_i\}},
\label{zc}
\end{equation}
which is essentially driven by the statistics of the minimum strength.
This is our general result.
To ensure that all node strengths are replicated satisfactorily in a typical single realization of the model, we need $z>z_c$. 
We are particularly interested in determining whether, in the realistic sparse regime $z=z_\mathrm{sparse}$ given by Eq.~\eqref{sparse}, the network is reconstructable, i.e. whether
\begin{equation}
z_\mathrm{sparse}>z_c.
\label{condition}
\end{equation}
Note that the above requirement, combined with the necessary sparsity condition in Eq.~\eqref{vanish}, implies another necessary condition:
\begin{equation}
z_c\to 0^+\quad \mathrm{for} \quad N\to+\infty.
\label{cvanish}
\end{equation}
In what follows, we are going to consider specific theoretical and empirical cases, corresponding to different distributions $f(s^*)$ of the node strengths.

\section{Specific cases\label{sec:cases}}
\subsection{Homogeneous networks\label{homo}}
As a first, trivial example, we consider the case of equal empirical strengths $s^*_i=1$ for all $i$, i.e. $f(s^*)=\delta(s^*-1)$. Inserting this specification into Eq.~\eqref{p}, it is clear that the underlying binary network reduces to an Erd\H{o}s-R\'enyi (ER) random graph with homogeneous connection probability \begin{equation}
p_{ij}(z)=\frac{z}{1+z}\equiv p(z)\qquad \forall i,j.
\end{equation}
Equation~\eqref{w} indicates that, when realized, each link gets the same conditional weight $w_{ij}=N^{-1}p^{-1}(z)=(1+z)/(Nz)$, so that the expected unconditional weight in Eq.~\eqref{expw} is $\langle w_{ij}\rangle=N^{-1}$ for each pair of nodes $i,j$.

Now, since in this particular case $p(z)$ coincides with the expected link density $d(z)$, Eq.~\eqref{sparse} implies that, to have a sparse network, we need $z=z_\mathrm{sparse}$ with
\begin{equation}
z_\mathrm{sparse}\simeq \frac{k}{N},\quad k>1
\label{sparse_ER}
\end{equation}
so that $d(z_\mathrm{sparse})=\frac{z_\mathrm{sparse}}{1+z_\mathrm{sparse}}\simeq z_\mathrm{sparse}=k/N$ as required.
Equation~\eqref{sparse_ER} sets the specific way in which the sparsity condition in Eq.~\eqref{vanish} is realized in this completely homogeneous case.
Note that the corresponding relative fluctuations of the strength are asymptotically constant:
\begin{equation}
    \delta_i(z_\mathrm{sparse})=\frac{1}{\sqrt{Nz_\mathrm{sparse}\, s^*_i}}\simeq\frac{1}{\sqrt{k\, s^*_i}}=\frac{1}{\sqrt{k}}\qquad \forall i
    \label{1/k}
\end{equation}
and hence do not vanish in the sparse case.
The critical value $z_c$ for the reconstructability found using Eq.~\eqref{zc} is 
\begin{equation}
z_c=\frac{1}{c}\,N^{\beta-1},\quad 0<\beta<1,
\label{zc_ER}
\end{equation}
where we have enforced the additional requirement $\beta<1$ to realize the necessary condition given in Eq.~\eqref{cvanish}.
Indeed, the `canonical' case $\beta=1$ (relative fluctuations vanishing like $1/\sqrt{N}$) would yield a finite threshold $z_c=c^{-1}>1$ and a finite critical density $d(z_c)=p_c=(1+c)^{-1}>1/2$. In this case, in the reconstructability phase $z>z_c$ the network would necessarily be dense.
Similarly, the case $\beta>1$ would yield $z_c=+\infty$ and $p_c=1$, so asymptotically the network would be a complete graph. 

Therefore, in the only possible case $0<\beta<1$ for the sparse regime, when $N$ is large we have $z_c\ll 1$ and hence $p_c\equiv p(z_c)=z_c/(1+z_c)\simeq z_c= c^{-1}N^{\beta-1}$. Importantly, $p_c$ vanishes more slowly than the critical threshold $p_\mathrm{gcc}\simeq 1/N$ associated with the onset of the giant connected component (${gcc}$) in the ER graph and also more slowly than the critical threshold $p_\mathrm{ucc}\simeq\log N/N$ associated with the onset of an overall connectivity in the ER graph (unique connected component, ${ucc}$). This clarifies that the reconstructability transition considered here is different from both transitions in the underlying ER model.
In particular, to ensure reconstructability $z$ need be larger than the values required for the entire network to have a giant and even a unique connected component. 

The above results suggest that, at least when all strengths are (nearly) equal to each other, one should necessarily set $0<\beta<1$ (hence deviating from the canonical choice for the scaling of the relative fluctuations) if the network has to simultaneously be reconstructable and have a vanishing density $d(z)\to 0$. 
In any case, $\beta>0$ implies that $z_c$ in Eq.~\eqref{zc_ER} is asymptotically always much larger than the value $z_\mathrm{sparse}$ in Eq.~\eqref{sparse_ER}, which is required to have a realistically sparse graph with density scaling as
in Eq.~\eqref{sparse}. 
Therefore the condition in Eq.~\eqref{condition} is always violated, except possibly for small values of $N$.
This means that real-world networks, if they were simultaneously large, sparse and homogeneous, could not be reconstructed with the approach considered here. 
Indeed we see that only for dense and homogeneous networks, i.e. finite $d(z)$, it is possible to achieve the reconstructability of the network, with $\beta=1$. Finally, $\beta>1$ is uninteresting as it leads to a fully connected network. 

\subsection{Core-periphery structure\label{core}}
We now consider a less trivial setting where we introduce some heterogeneity in the network in the form of a core-periphery structure, where nodes in the core have a larger strength and nodes in the periphery have a smaller strength. 
The presence of a core-periphery structure has been documented in real financial networks
\cite{alves2013structure,barucca2016disentangling,fricke2015core,van2014finding}. 
For simplicity we assume that all nodes in the core are still homogeneous, i.e. they all have the same value $s_c^*$ of the expected strength, and the same goes for all nodes in the periphery, i.e. they all have the same expected strength $s_p^*$, with clearly $s_p^*<s_c^*$.
We place a fraction $q$ of the $N$ nodes in the core, i.e. the core has $N_c=qN$ nodes and the periphery has $N_p=(1-q)N$ nodes. Then the distribution of empirical strengths is
\begin{equation}
f(s^*)=(1-q)\,\delta(s^*-s^*_p)+q\,\delta(s^*-s^*_c)    
\end{equation}
and, setting $\overline{s^*}=1$, we get 
\begin{equation}
\overline{s^*}=(1-q)\,s^*_p+q\,s^*_c\equiv 1
\end{equation}
where $s^*_p<1<s^*_c$.
Inverting,
\begin{equation}
q=\frac{1-s^*_p}{s^*_c-s^*_p}.
\label{q}
\end{equation}
Since $\textrm{min}_i\{s^*_i\}=s^*_p$, Eq.~\eqref{zc} indicates that the critical density is achieved by the condition
\begin{equation}
z_c=\frac{1}{c\,s^*_p}\,N^{\beta-1},\quad \beta,c>0
\label{zc_cp}
\end{equation}
whose asymptotic behaviour essentially depends on how $s^*_p$ is chosen to scale as $N$ grows.

As in the previous example, we need to compare the above $z_c$ with the value $z_\mathrm{sparse}$ given by Eq.~\eqref{sparse}. To this end, we first compute the expected density defined in Eq.~\eqref{density} as
\begin{equation}
d(z)=d_{cc}(z)+d_{pp}(z)+d_{cp}(z),
\end{equation}
where $d_{cc}(z)$, $d_{pp}(z)$ and $d_{cp}(z)$ represent the core-core, periphery-periphery and core-periphery contributions given by
\begin{equation}
d_{cc}(z)=\frac{N_c(N_c-1)}{N(N-1)}\frac{z\,(s^*_c)^2}{1+z\,(s^*_c)^2},
\label{dcc}
\end{equation}
\begin{equation}
d_{pp}(z)=\frac{N_p(N_p-1)}{N(N-1)}\frac{z\,(s^*_p)^2}{1+z\,(s^*_p)^2},
\label{dpp}
\end{equation}
\begin{equation}
d_{cp}(z)=\frac{2N_c\,N_p}{N(N-1)}\frac{z\,s^*_c\,s^*_p}{1+z\,s^*_c\,s^*_p},
\label{dcp}
\end{equation}
respectively.
Next, in order to look for $z_\mathrm{sparse}$ given by Eq.~\eqref{sparse}, we need to specify how $N_c$, $N_p$, $s^*_c$ and $s^*_p$ scale with $N$.

If both $N_c$ and $N_p$ grow linearly in $N$ and if both $s^*_c$ and $s^*_p$ remain finite (with $s^*_p<1<s^*_c$), then it is easy to see that the three terms in Eqs.~\eqref{dcc},~\eqref{dpp} and~\eqref{dcp} are all of the same order and the implied $z_\mathrm{sparse}$ is inversely proportional to $N$, as in the previous example. Again, this means $z_\mathrm{sparse}<z_c$ and the network is unreconstructable in the sparse regime. 

To better exploit the potential of the core-periphery model, we therefore consider a highly concentrated (or `condensed') case where, as $N$ increases, only a finite number $N_c$ of nodes remain in the core, while the size $N_p=N-N_c$ of the periphery grows extensively in $N$. We also assume that the strength of peripheral nodes decreases with $N$ as 
\begin{equation}
s^*_p\simeq \alpha\, N^{-\eta},\quad \alpha,\eta> 0.
\end{equation}
Since $q= N_c/N$, Eq.~\eqref{q} implies that asymptotically 
\begin{equation}
s^*_c\simeq q^{-1}=N/N_c.
\end{equation}
Inserted into Eqs.~\eqref{dcc},~\eqref{dpp} and~\eqref{dcp}, the above specifications imply that 
$d_{cc}(z)$ is subleading with respect to $d_{pp}(z)$ and $d_{cp}(z)$. Therefore we can look for $z_\mathrm{sparse}$ by setting $d_{pp}(z_\mathrm{sparse})+d_{cp}(z_\mathrm{sparse})\simeq k/N$.
Taken separately, the requirement $d_{pp}(z_\mathrm{sparse})\simeq k_1 N^{-1}$ for some $k_1>0$ implies  
\begin{equation}
    z_\mathrm{sparse}\simeq \frac{k_1}{N(s^*_p)^2}\simeq\frac{k_1}{\alpha^2}\, N^{2\eta-1},\quad0<\eta<{1}/{2},
    \label{zsparse_cp}
\end{equation}
where we have enforced the extra condition $\eta<1/2$ to ensure that the exponent $2\eta-1$ of $N$ is negative, in order to meet the necessary condition given by Eq.~\eqref{vanish}.
When inserted into $d_{cp}(z)$, Eq.~\eqref{zsparse_cp} implies
$d_{cp}(z_\mathrm{sparse})\simeq k_2/N$ with $k_2=2N_c$, showing that $d_{pp}(z_\mathrm{sparse})$ and $d_{cp}(z_\mathrm{sparse})$ are of the same order and both contribute to the desired scaling $d(z_\mathrm{sparse})\simeq k/N$ with $k=k_1+k_2$.

The above scaling of $z_\mathrm{sparse}$ has to be compared with the reconstructability threshold $z_c$ in Eq.~\eqref{zc_cp} which, for the chosen behaviour of $s^*_p$, reads
\begin{equation}
z_c=\frac{1}{c\,\alpha}\,N^{\beta-1+\eta},\qquad\alpha,\beta,c>0, \quad 0<\eta<{1}/{2}.
\label{zc_cp2}
\end{equation}
Comparing Eqs.~\eqref{zsparse_cp} and~\eqref{zc_cp2}, we see that the reconstructability condition $z_\mathrm{sparse}>z_c$ is met asymptotically (for large $N$) if 
\begin{equation}
    0<\beta<\eta,
    \label{disegual}
\end{equation}
while it is asymptotically violated if $0<\eta<\beta$.
Indeed the largest relative fluctuations in the sparse regime, following Eq.~\eqref{delta_zmax}, are given by
\begin{equation}
\max_{i}\{\delta_i(z_\mathrm{sparse})\}=\frac{1}{\sqrt{N\,z_\mathrm{sparse}\,s^*_p}}\simeq\sqrt{\frac{s^*_p}{k_1}}\simeq \sqrt{\frac{\alpha}{k_1}}\, N^{-\eta/2}
\label{delta_zmaxcp}
\end{equation}
and they remain within the critical value given by Eq.~\eqref{delta_zc} (with the identification $c=\alpha/k_1$) if the condition in Eq.~\eqref{disegual} is met. So, for any chosen $\beta>0$, there is a critical exponent $\eta_\beta\equiv\beta$ such that for $\eta>\eta_\beta$ the network is reconstructable while for for $\eta<\eta_\beta$ the network is unreconstructable.
Since $0<\eta<{1}/{2}$, we see that reconstructability is possible only if $0<\beta<1/2$. So, also in this case, the canonical decay $\beta=1$ of the relative fluctuations cannot be achieved in the sparse regime.

The core-periphery model considered here is a simple, yet important example showing that a sufficiently heterogeneous network structure, by acting on the statistics of the minimum strength, can imply the presence of two phases (reconstructability and unreconstructability) in the sparse regime.
Contrary to what naive intuition may suggest by looking at Eq.~\eqref{zc}, i.e. that a decreasing (with $N$) value of $\textrm{min}_i\{s^*_i\}$ could make the reconstructability only more difficult by increasing the value of $z_c$, actually the net effect might be the opposite if at the same time $z_\mathrm{sparse}$ increases as well, sufficiently fast. This circumstance, which is impossible in the completely homogeneous case discussed in Sec.~\ref{homo}, is precisely what happens in Eq.~\eqref{zsparse_cp} due to the presence of $(s^*_p)^2$, rather than $s^*_p$ itself, in the denominator. 

\subsection{Arbitrary strength distribution\label{arbitrary}}
We now generalize the calculation of the critical reconstructability threshold $z_c$ to the case of an arbitrary empirical strength distribution. 
To calculate $\textrm{min}_{i=1,N}\{s^*_i\}$ as a function of $N$, we assume that the $N$ values $\{s^*_i\}_{i=1}^N$ of the empirical strength are realized via sampling $N$ times (in an i.i.d. manner) from some probability density function $f(s)$.
Then, a simple argument from Extreme Value Theory indicates that the typical value of the realized minimum strength $s^*_\mathrm{min}(N)=\textrm{min}_{i=1,N}\{s^*_i\}$ is such that the expected number of nodes with strength $s^*\le s^*_\mathrm{min}$ is of order one, or equivalently
\begin{equation}
\frac{1}{N}\simeq\int_0^{s^*_\textrm{min}(N)}f(s)\,ds.
\label{integralmin}
\end{equation}
For a given choice of $f(s)$, inverting the above equation produces the sought-for scaling of $s^*_\textrm{min}(N)$ with $N$.

We are mainly interested in the behaviour of $f(s)$ for low values $s\to 0^+$ of the strength, because that is the behaviour that determines the statistics of the minimum strength and the behaviour of the integral in Eq.~\eqref{integralmin}.
In particular we consider the following behaviour
\begin{equation}
f(s)\to_{s\to0^+}a\, s^{\psi-1},\quad a,\psi>0.
\end{equation}
Inserting this into Eq.~\eqref{integralmin} we get
\begin{equation}
\frac{1}{N}\simeq a\int_0^{s_\textrm{min}(N)}s^{\psi-1}\,ds=\frac{a}{\psi}\,s_\textrm{min}^\psi(N)\label{integralmin2}
\end{equation}
and, inverting,
\begin{equation}
s_\textrm{min}(N)\simeq \left(\frac{\psi}{a}\right)^{1/\psi}N^{-1/\psi}
\label{smin}
\end{equation}
This implies 
\begin{equation}
z_c=\frac{(a/\psi)^{1/\psi}}{c}\,N^{\beta-1+1/\psi}
\end{equation}
where, to realize the  condition in Eq.~\eqref{cvanish}, we need $\beta-1+1/\psi<0$ or equivalently
\begin{equation}
\psi>(1-\beta)^{-1}.
\label{psipsi}
\end{equation}

A comparison with Eq.~\eqref{zc_cp2} highlights that $1/\psi$ has an effect on $z_c$ similar to the one that $\eta$ has in the core-periphery model.
Indeed, if in general we have a scaling of the form \begin{equation}
    s_\textrm{min}(N)\simeq m\,N^{-\mu}\qquad m>0,\,\mu\ge 0,
\label{mu}
\end{equation}
then Eq.~\eqref{zc} implies
\begin{equation}
    z_c\simeq \frac{1}{c\,m}\,N^{\beta-1+\mu}
\end{equation}
where, to realize the condition in Eq.~\eqref{cvanish}, we need 
\begin{equation}
    0<\mu<1-\beta.
\end{equation}
In the examples considered so far, $\mu\equiv 0$ (with $m\equiv 1$) for the homogeneous case discussed in Sec.~\ref{homo} (confirming that the above inequality cannot be realized), while $\mu\equiv \eta$ with $0<\eta<1/2$ for the core-periphery model discussed in Sec.~\ref{core} (for which we need $0<\beta<1/2$), and finally $\mu\equiv1/\psi$ for the general case discussed in this Section, for which Eq.~\eqref{psipsi} has to hold.
In order to check for reconstructability, one should of course calculate also $z_\mathrm{sparse}$, which requires the full knowledge of $f(s)$ and has to be evaluated for the specific case at hand.

\section{The role of induced self-loops\label{sec:loops}}
As anticipated, we now show that, even though the probabilistic reconstruction model considered here generates self-loops, the role of the latter in the reconstructability problem is negligible for the typical regimes of relevance for real-world networks.
We recall that, in our units such that $\overline{s^*}=1$, the total weight of all links in the empirical network is $W^*=\sum_{i=1}^N s^*_i/2=N/2$ and its expected value in the model replicates the empirical value perfectly, i.e. $\langle W\rangle=W^*$. However the model, differently from the real network, produces self-loops, each with an expected weight $\langle w_{ii}\rangle$ given by Eq.~\eqref{looppaccio}. The resulting expected total weight of self-loops can be calculated as  
\begin{equation}
\langle W_\mathrm{SL}\rangle=\sum_{i=1}^N \langle w_{ii}\rangle=\sum_{i=1}^N \frac{(s^*_i)^2}{2W^*}=\frac{N\overline{s^{*2}}}{2W^*}={\overline{s^{*2}}}
\label{looppacciotot}
\end{equation}
where $\overline{s^{*2}}=\sum_{i=1}^N (s^*_i)^2/N$ is the empirical second moment of the node strengths.
If we require that the weight of all self-loops is negligible with respect to the expected weight $\langle W\rangle$ of all links (including self-loops), we should impose
\begin{equation}
\frac{\langle W_\mathrm{SL}\rangle}{\langle W\rangle}=\frac{2\overline{s^{*2}}}{N}=o(1)\quad\implies\quad \overline{s^{*2}}=o(N)
\label{eq:condition}
\end{equation}
Note that Eq.~\eqref{eq:condition} essentially sets a bound on the second moment, hence on the right tail, of the strength distribution $f(s)$. This should be combined with the previous bounds on the left tail.

We now show that the above condition is satisfied in the regimes that are typical for real-world networks.
In particular, if we assume realistic strength distributions that are observed in empirical financial networks, we should mainly consider log-normal distributions and distributions with a power-law tail decay $f(s)\simeq b\, s^{-\gamma}$ (with $b>0$) for large $s$.
If the second moment of $f(s)$ is finite (as for the log-normal distribution and the power-law distribution with $\gamma>3$), we can automatically conclude that Eq.~\eqref{eq:condition} is verified. 
If the second moment of $f(s)$ diverges (as for the power-law distribution with $\gamma\le 3$), we should more carefully look at how fast this occurs as $N$ grows and check whether Eq.~\eqref{eq:condition} is still satisfied.
We can do so using an Extreme Value Theory argument analogous to the one in Eq.~\eqref{integralmin}: we can first estimate how the realized maximum strength $s^*_\textrm{max}(N)$ (out of an i.i.d. sample of $N$ values) scales as a function of $N$, and then use $s^*_\textrm{max}(N)$ to establish how the realized empirical second moment $\overline{s^{*2}}$ scales with $N$.
Considering the non-trivial case $f(s)\simeq b\,s^{-\gamma}$ for large $s$, we estimate $s^*_\textrm{max}(N)$ in analogy with Eq.~\eqref{integralmin} as
\begin{equation}
\frac{1}{N}\simeq\int_{s^*_\textrm{max}(N)}^{+\infty}f(s)\,ds\simeq b\int_{s^*_\textrm{max}(N)}^{+\infty}s^{-\gamma}\,ds,
\label{integralmax}
\end{equation}
which leads to (note that $\gamma-1>0$)
\begin{equation}
s^*_\textrm{max}(N)\simeq \left(\frac{b}{\gamma-1}\right)^{\frac{1}{\gamma-1}} N^{\frac{1}{\gamma-1}}.
\end{equation}
We can then estimate the scaling of $\overline{s^{*2}}(N)$ as follows:
\begin{eqnarray}
\overline{s^{*2}}(N)&=&\int_0^{s^*_\textrm{max}(N)}f(s)\, s^2\, ds\nonumber\\
&\simeq&b\int_0^{s^*_\textrm{max}(N)} s^{2-\gamma}\, ds\nonumber \\
&\simeq&\frac{b}{\gamma-3}\, {[s^*_\textrm{max}(N)]}^{3-\gamma}\nonumber\\
&\simeq&\frac{b}{\gamma-3}\,
\left(\frac{b}{\gamma-1}\right)^{\frac{3-\gamma}{\gamma-1}}\, N^{\frac{3-\gamma}{\gamma-1}}.
\label{mecojoni}
\end{eqnarray}
To ensure that self-loops contribute only negligibly as stated in Eq.~\eqref{eq:condition}, we need to require $(3-\gamma)/(\gamma-1)<1$, which simply boils down to $\gamma>2$.
Importantly, this means that empirical strength distributions that have an asymptotically diverging second moment are still viable, provided they have an asymptotically finite first moment. This is indeed what is observed in most real-world financial networks, where the strength distribution typically has either a log-normal form or a power-law tail with an exponent in the range $2<\gamma<3$. So all our calculations in the previous sections are justified in the regimes of relevance for real-world networks.

The same results can be derived more rigorously by analysing the large $N$ convergence of the probability distribution of the random variable $W_\mathrm{SL}$ (whose possible realized values will be denoted as $w_\mathrm{SL}$).
Let us assume that the probability distribution $f(s)$ of node strengths does not depend on $N$, in agreement with the assumption of sparse networks with $2W^*= N$.
Let us consider the PDF $g(u)$ of the variable $u=s^2$:
\[g(u)=\frac{f(\sqrt{u})}{2\sqrt{u}}\,.\]
The PDF $p(w_\mathrm{SL})$ of $W_\mathrm{SL}$ is related to $g(u)$ by
\[p(w_\mathrm{SL})=\int_{0}^{\infty}\cdots\int_{0}^{\infty}\left[\prod_{i=1}^Ndu_i\,g(u_i)\right]\delta\left(w_\mathrm{SL}-\sum_{i=1}^N \frac{u_i}{N}\right)\]
Consequently, its characteristic function is
\begin{equation}
\hat p (t)=\int_{0}^{\infty}dw_\mathrm{SL} \,p(w_\mathrm{SL})e^{-tw_\mathrm{SL}}=\left[\int_{0}^{\infty}du\,g(u)e^{-tu/N}\right]^N\,.
\label{eq1}
\end{equation}
We have to distinguish two fundamental cases: (i) $u$ has a finite mean value (i.e. $s$ has a finite variance); (ii) $u$ has diverging mean value (i.e. $s$ has an infinite variance).

\begin{itemize}
\item $\overline u=\overline {s^2}<\infty$: in this case at small enough $t$ we have
\begin{equation}
\int_{0}^{\infty}du\,g(u)e^{-tu/N}=1-\frac{\overline u}{N}t+o(t/N)\,.
\end{equation}
Using this expression in Eq.~(\ref{eq1}) and taking the large $N$ limit we get
\begin{equation}
\hat p (t)=e^{-\overline u t}\,.
\label{eq2}
\end{equation}
which is the Laplace transform of $p(w_\mathrm{SL})=\delta(w_\mathrm{SL}-\overline u)$.
Using Eq.~\eqref{looppacciotot} we can therefore conclude that in this case the relative weight of self-loops is $\langle W_\mathrm{SL}\rangle/\langle W\rangle=O(1/N)$, in accordance with Eq.~\eqref{eq:condition}. This confirms that log-normal distributions or power-law distributions with finite variance ($\gamma>3$) are always acceptable.\\

\item $\overline u=\overline {s^2}\to\infty$: in this case we consider again the case $f(s)\simeq b\, s^{-\gamma}$ at large $s$ with $1<\gamma\le 3$ (note that $\gamma>1$ is required in order to guarantee the integrability of the PDF). Consequently $g(u)\simeq (b/2)\, u^{-(\gamma+1)/2}$ at large $u$ (note that $\gamma>1$ implies $(\gamma+1)/2>1$ too).
By the properties of the Laplace transform we now have that, at small $t$,
\begin{equation}
\int_{0}^{\infty}du\,g(u)e^{-tu/N}=1-A\frac{|t|^{(\gamma-1)/2}}{N^{(\gamma-1)/2}}+o[(|t|/N)^{(\gamma-1)/2}]
\end{equation}
for some finite $A$.
If we now use this expression in Eq.~(\ref{eq1}) at large but finite $N$ we have 
\begin{equation}
\hat p(t)\simeq e^{-AN^{(3-\gamma)/2}|t|^{(\gamma-1)/2}}\,
\label{eq3}
\end{equation}
which implies the typical value $W_\mathrm{SL}\sim N^{(3-\gamma)/(\gamma-1)}$. 
Recalling Eq.~\eqref{looppacciotot}, this result coincides with Eq.~\eqref{mecojoni}.
Indeed Eq.~(\ref{eq3}) implies $p(w_\mathrm{SL})\sim w_\mathrm{SL}^{-(\gamma+1)/2}$ for $w_\mathrm{SL}\gg N^{(3-\gamma)/(\gamma-1)}$ and, in order to have a total measure of order 1 for the variable $W_\mathrm{SL}$, we have to go up to values of order $N^{(3-\gamma)/(\gamma-1)}$.
From Eq.~\eqref{eq:condition} we see that the weight of self-loops can be neglected if $(3-\gamma)/(\gamma-1)<1$ i.e. $2<\gamma<3$. This coincides with the `realistic' regime identified above.

\end{itemize}

We can summarize the discussion above by saying that self-loops can be neglected with respect to the rest of the connections if the empirical strength distribution has a finite variance (as in log-normal distributions or power-law distributions with $\gamma>3$) or if it has a power-law tail with $2<\gamma<3$ (such that the mean is finite but the variance diverges). Realistic situations typically fall into one of those two cases. The next Section provides an empirical example that confirms this picture.

\section{Reconstructed networks from Bankscope Data\label{sec:data}}
\subsection{Dataset}\label{data_description}
The Bureau Van Dijk \emph{Bankscope} database contains information on banks' balance sheets and aggregate exposures. Our dataset consists of a subset of $N = 119$ anonymized European banks that were publicly traded between 2006 and 2013~\cite{battiston2016leveraging}. For each of the $N$ banks and each year $t$ in the data, we have access to the yearly values of total interbank assets, total interbank liabilities, total assets, total liabilities, and equity.

\begin{figure}[t!]
\centering
\includegraphics[width=1\textwidth]{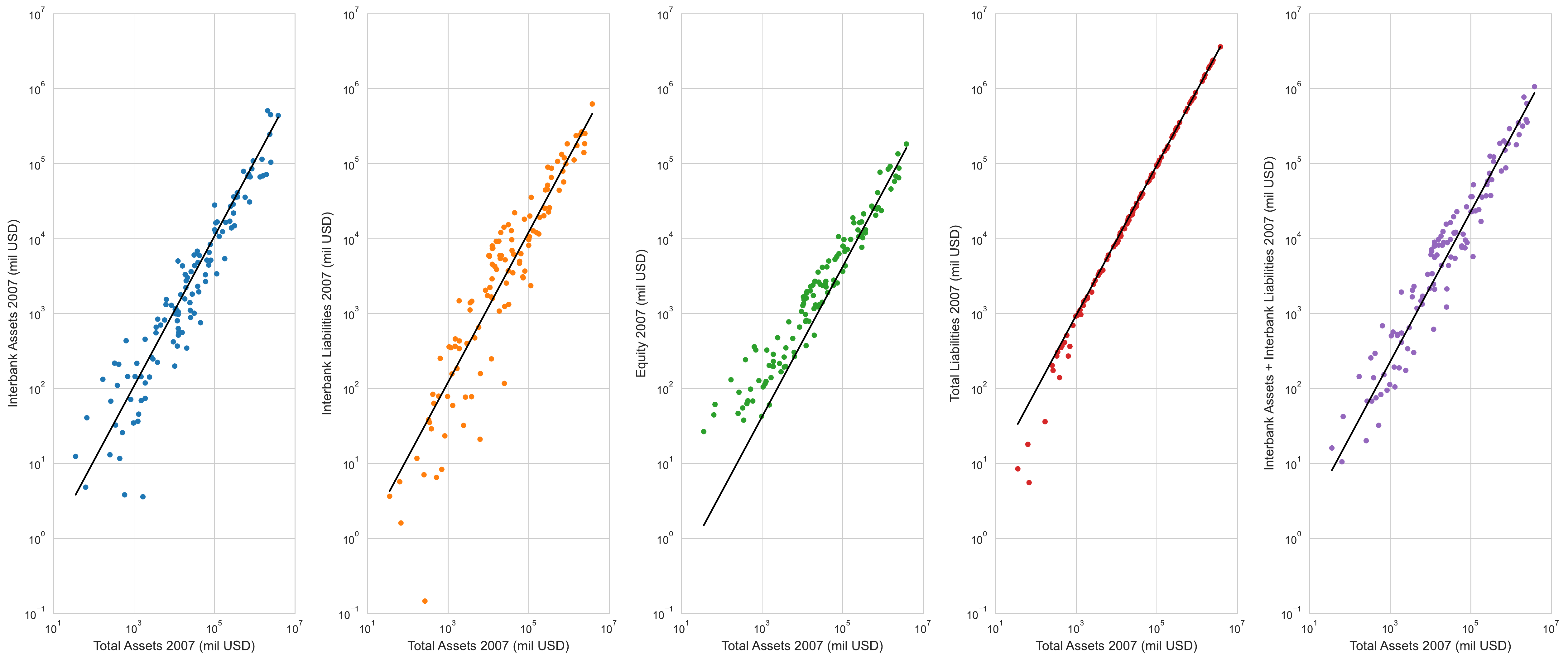}
\caption{Scatter plot and simple linear regression of the type $y_{i,k}(t)=a_k(t)\cdot x_i(t)$ of the Interbank Assets ($k=1$), Interbank Liabilities ($k=2$), Equity ($k=3$), Total Liabilities ($k=4$), and Interbank Assets + Liabilities ($k=5$) of each bank $i$ ($i=1,119$) in year $t=2007$, versus the total assets $x_i(t)$ of the same bank in the same year. Data from Bankscope, reported in mil USD.}
\label{fig:LR_bankscope}
\end{figure}

To characterize the data, we start by noticing that all these bank-specific variables are strongly linearly correlated, as the plots in Fig.~\ref{fig:LR_bankscope} illustrate for year 2007.
This observation holds for all the years in the data. Indeed, in Table~\ref{tab:cof} we report, for each year $t$ in the data, the fitted proportionality coefficient $a_k(t)$ and the corresponding coefficient of determination $R^2_k(t)$ of a simple linear regression of the type $y_{i,k}(t)=a_k(t)\cdot x_i(t)$, where the independent variable $x_i(t)$ is always the total assets of bank $i$ in year $t$ and the dependent variable $y_{i,k}(t)$ is the Interbank Assets ($k=1$), Interbank Liabilities ($k=2$), Equity ($k=3$), Total Liabilities ($k=4$), and Interbank Assets+Liabilities ($k=5$) of the same bank in the same year.

\begin{table}[t!]
    \centering
\begin{tabular}{|c|l|cccccccc|}
\toprule
\multicolumn{2}{|c|}{{Year $t$}} &    ~2006~ &    2007~ &    2008~ &    2009~ &    2010~ &    2011~ &    2012~ &   2013~ \\
\midrule
&Interbank Assets                         &  0.11 &  0.11 &  0.06 &  0.06 &  0.06 &  0.06 &  0.05 &  0.06 \\
&Interbank Liabilities                    &  0.12 &  0.12 &  0.10 &  0.08 &  0.07 &  0.07 &  0.06 &  0.05 \\
$a_k(t)$ &Equity                                   &  0.04 &  0.04 &  0.03 &  0.05 &  0.05 &  0.05 &  0.05 &  0.05 \\
&Total Liabilities                        &  0.96 &  0.96 &  0.97 &  0.95 &  0.95 &  0.95 &  0.95 &  0.95 \\
&Interbank Assets+Liabilities~ &  0.23 &  0.23 &  0.16 &  0.15 &  0.13 &  0.13 &  0.11 &  0.11 \\
\midrule
&Interbank Assets                         &  0.69 &  0.74 &  0.46 &  0.53 &  0.47 &  0.46 &  0.41 &  0.44 \\
&Interbank Liabilities                    &  0.87 &  0.88 &  0.86 &  0.90 &  0.86 &  0.80 &  0.71 &  0.65 \\
$R_k^2(t)$ &Equity                                   &  0.87 &  0.89 &  0.88 &  0.92 &  0.93 &  0.93 &  0.92 &  0.93 \\
&Total Liabilities                        &  0.99 &  0.99 &  0.99 &  0.99 &  0.99 &  0.99 &  0.99 &  0.99 \\
&Interbank Assets+Liabilities~ &  0.85 &  0.89 &  0.78 &  0.83 &  0.73 &  0.69 &  0.62 &  0.59 \\
\bottomrule
\end{tabular}
\vskip 5pt
\caption{Coefficient $a_k(t)$ (top) and oefficient of determination $R_k^2(t)$ (bottom) of a simple linear regression of the type $y_{i,k}(t)=a_k(t)\cdot x_i(t)$, where $x_i(t)$ is the total assets of bank $i$ in year $t$ (for $t=2006,\dots,2013$) and $y_{i,k}(t)$ is the Interbank Assets ($k=1$), Interbank Liabilities ($k=2$), Equity ($k=3$), Total Liabilities  ($k=4$), and Interbank Assets + Liabilities  ($k=5$) of the same bank in the same year. Data from Bankscope.}
\label{tab:cof}
\end{table}

The approximate linearity of all the quantities allows us to proceed with an undirected description of the data, in line with our discussion so far. We define the strength of node $i$ as
\begin{equation}
    s_i^*=A_i+L_i
    \label{eq:a+l}
\end{equation}
where $A_i$ and $L_i$ are the total interbank assets and interbank liabilities of bank $i$, respectively. From now on, we limit ourselves to year 2007 and rescale the yearly strengths to the average strength so that $\overline{s^*}=1$, in line with our choice of units so far.

\begin{figure}[t]
     \centering
     \begin{subfigure}[b]{0.45\textwidth}
         \centering
         \includegraphics[width=\textwidth]{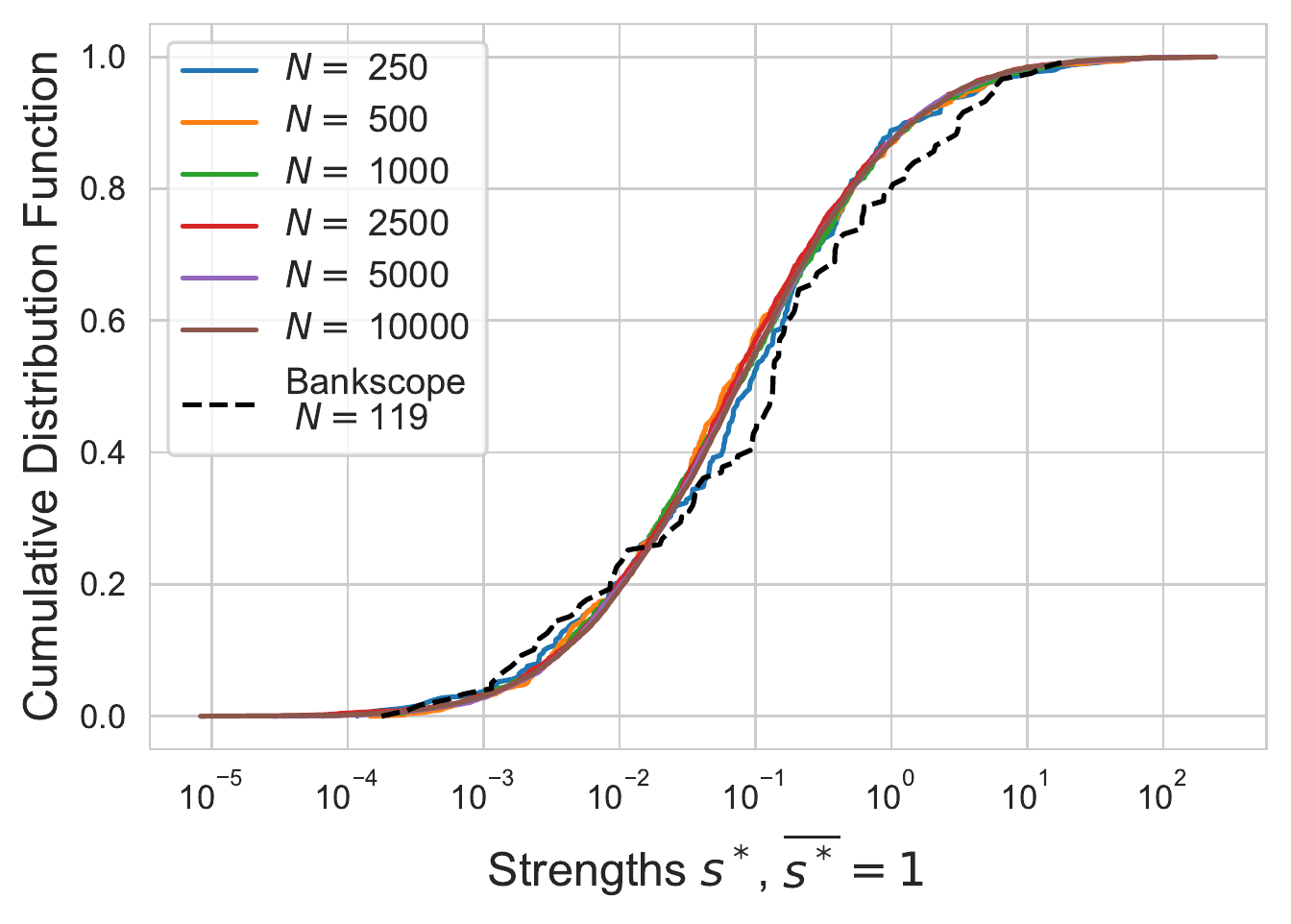}
         \caption{\label{fig:suba}}
     \end{subfigure}
     \hfill
     \begin{subfigure}[b]{0.475\textwidth}
         \centering
         \includegraphics[width=\textwidth]{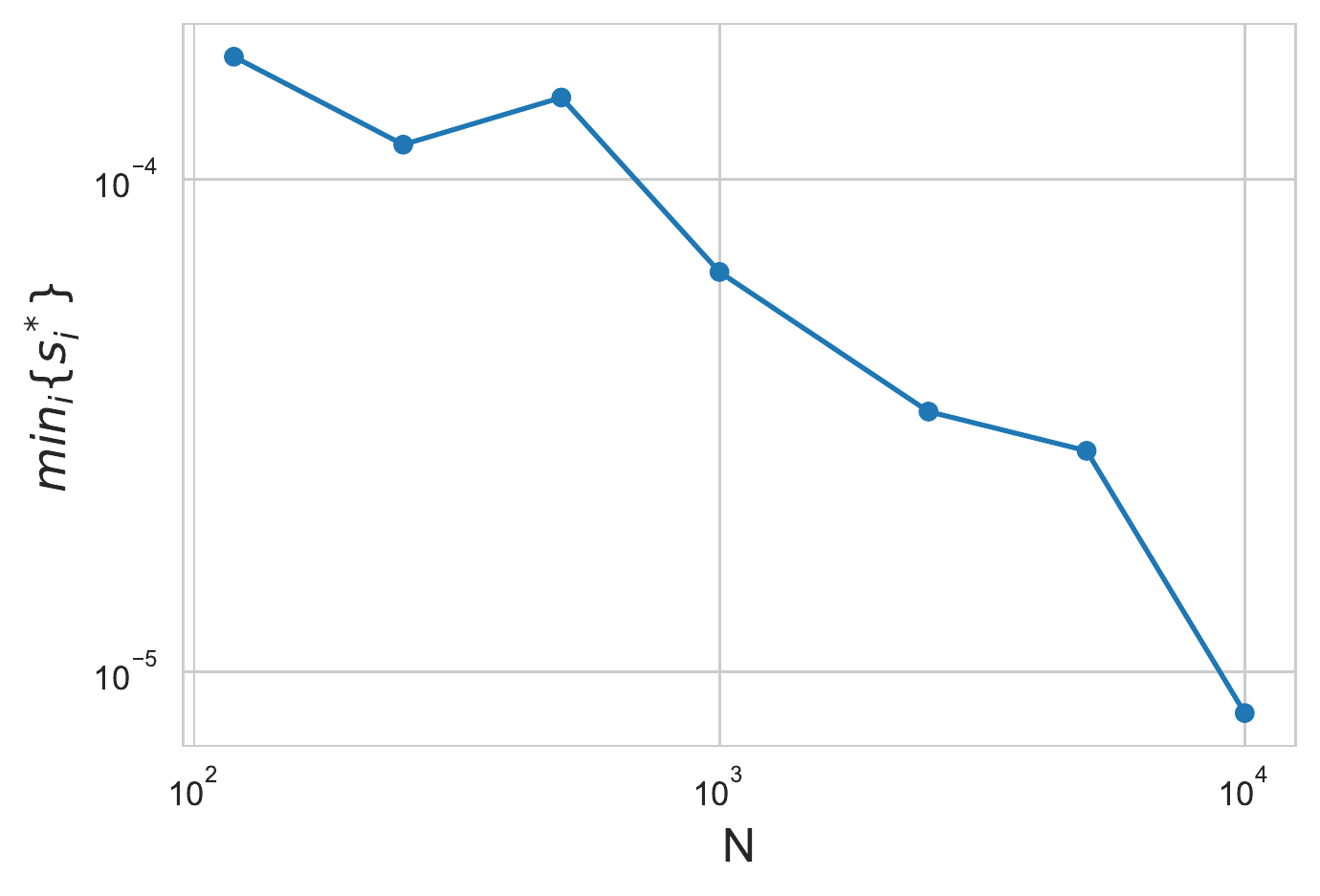}
         \caption{\label{fig:subb}}
     \end{subfigure}
        \caption{Empirical and synthetic data, using the 2007 Bankscope dataset as a reference. (a) Cumulative distribution function of the rescaled node strengths for the Bankscope data ($N=119$) and for the synthetic data (for increasing values of $N$) drawn from a  log-normal distribution fitted to empirical data (with $\sigma=2.28$ and $\mu=-\sigma^2/2$). (b) Minimum realized strength $s^*_\mathrm{min}(N)$ as a function of the number $N$ of nodes. Again, the point at $N=119$ is the Bankscope dataset.\label{fig:Bankscope}}
\end{figure}

The set of empirical strengths $\{s^*_i\}_{i=1}^{119}$ represents the starting point of our analysis and its empirical cumulative distribution function (CDF) is shown in Fig.~\ref{fig:suba}. Since many of our results refer to the behavior of $z_c$ and $z_\mathrm{sparse}$ as $N$ grows, we also construct synthetic replicas of the dataset for increasing values of $N$. 
To achieve this, we first fit a log-normal distribution 
\begin{equation}
    f(s)={\frac {1}{s\,\sigma {\sqrt {2\pi }}}}\ \exp \left(-{\frac {\left(\ln \left(s\right)-\mu \right)^{2}}{2\sigma ^{2}}}\right),\quad s>0
\end{equation}
to the empirical distribution of the $N=119$ node strengths and then use it to sample any desired number $N$ of i.i.d. strength values $\{s^*_i\}_{i=1}^{N}$ from it. Note that the theoretical mean value of the log-normal distribution is 
\begin{equation}
\overline{s}=\int_0^{+\infty}f(s)\,s\,ds=\exp \left(\mu +{\frac {\sigma ^{2}}{2}}\right),
\end{equation}
therefore, to ensure an expected unit mean value $\overline{s^*}=1$ for the sampled strengths, we set $\mu\equiv-\sigma^2/2$, leaving out only the free parameter $\sigma$.
When fitted to the data, the latter gets the value $\sigma=2.28$.
The resulting theoretical CDF of the strengths is
\begin{equation}
    F(s)=\int_0^{s}f(x)\,dx={\frac {1}{2}}\left[1+\operatorname {erf} \left({\frac {\ln s+\sigma^2/2 }{\sigma {\sqrt {2}}}}\right)\right],\quad s>0.
    \label{CDF}
\end{equation}
Figure~\ref{fig:suba} shows the good agreement between the empirical CDF of the 119 empirical Bankscope strengths and the CDFs of the corresponding $N$ synthetic values (for increasing $N$) sampled from the fitted log-normal distribution with CDF given by Eq.~\eqref{CDF}.
Note that the accordance with the log-normal distribution automatically ensures that the induced self-loops in the network reconstruction method do not represent any problem, as discussed in Sec.~\ref{sec:loops}. We will confirm this result explicitly later on.
We also report, in Fig.~\ref{fig:subb}, the value of the realized minimum strength $s^*_\mathrm{min}(N)=\min_{i=1,N}\{s^*_i\}$ as a function of $N$. As in our examples considered in Secs.~\ref{core} and~\ref{arbitrary}, $s^*_\mathrm{min}(N)$ decreases with $N$, realizing the necessary condition given in Eq.~\eqref{vanish} to have a sparse network. The behaviour of $s^*_\mathrm{min}(N)$ will also affect the critical value $z_c$ given by Eq.~\eqref{zc} for the reconstructability of the network.

In our following analysis, we will also consider a completely homogeneous benchmark where, for a given set of $N$ banks, each bank $i$ is assigned exactly the same strength $s^*_i=1$, as in our discussion in Sec.~\ref{homo}. This benchmark (in which clearly $s^*_\mathrm{min}=1$ independently of $N$) will serve as a reference to emphasize the role of bank heterogeneity in the reconstructed networks.

\subsection{Network reconstruction}
We now apply the method described in Sec.~\ref{reconstruction} to the sets of empirical and synthetic strengths to test our theoretical results and check for possible transitions from reconstructability to unreconstructability. 
We are interested in the sparse regime where the link density is given by $d(z_\mathrm{sparse})\simeq k/N$ as in Eq.~\eqref{sparse}, where $k$ is the average node degree in the network.
For a given strength distribution, we explore various numbers of nodes, $N=\{119,250,500,1000,\dots,10000\}$, and two values of the average degree, $k=\{50,100\}$. For each pair of $N$ and $k$, we consider an ensemble of 1000 realizations of weighted undirected networks. We are interested in determining whether typical realizations produce the desired strengths with vanishing relative fluctuations, i.e., whether Eq.~\eqref{delta_zc} is verified.

\begin{figure}[t!]
\centering
\includegraphics[width=1\textwidth]{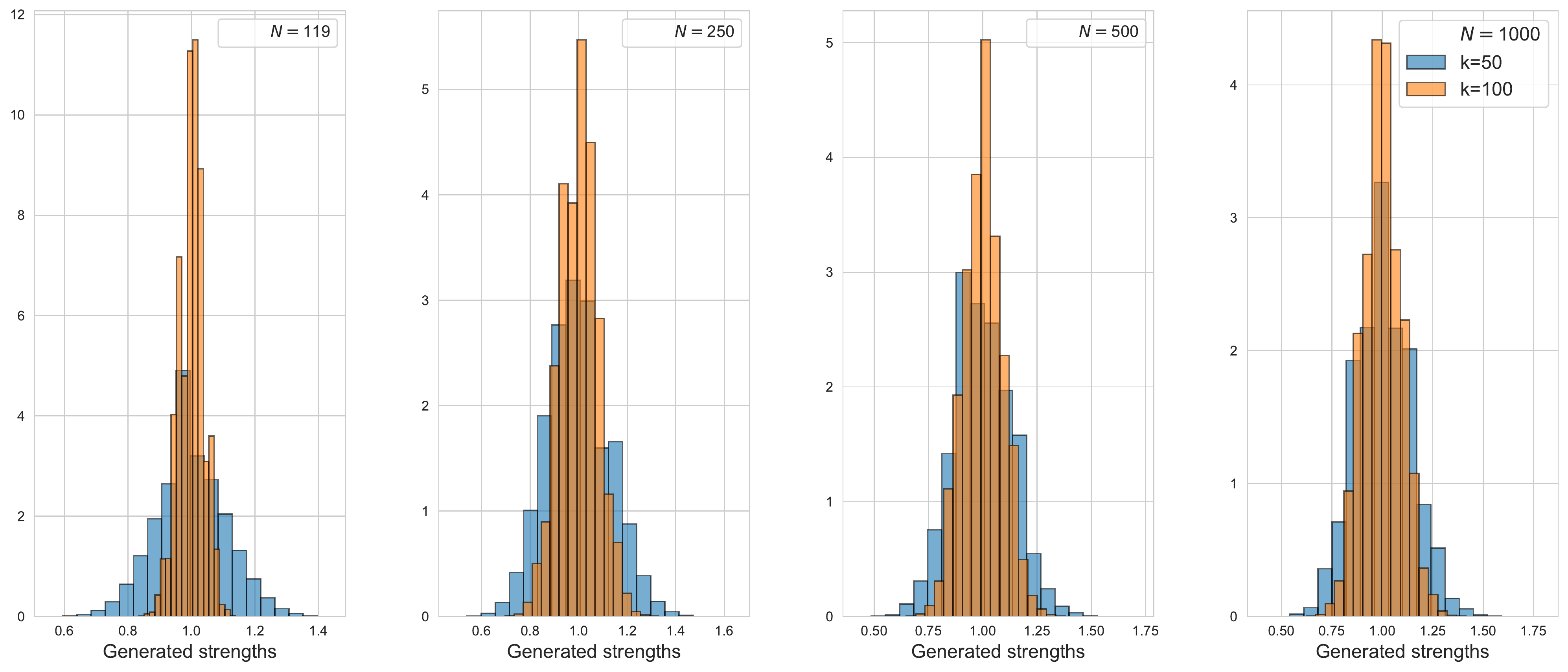}
\caption{Distribution of the reconstructed strengths in the case of identical nodes (homogeneous networks), across all the nodes and all the 1000 network realizations in the ensemble. The distribution is peaked around the expected value $\overline{s^*}=1$. Different colors correspond to different choices of the constant $k$, while each subplot corresponds to a different number $N$ of nodes.}
\label{fig:equal_realvsgen}
\end{figure}

\subsubsection{Identical strengths.}
We start from the homogeneous benchmark where all banks are identical and all strengths are therefore equal. As discussed in Sec.~\ref{homo}, the underlying binary network reduces to an Erd\H{o}s-R\'enyi random graph with homogeneous connection probability. 

In Fig.~\ref{fig:equal_realvsgen} we report, for different values of $N$ and $k$, the distribution of the realized strengths in the networks sampled from the ensemble, across all the $N$ nodes and all the 1000 network realizations. The results confirm a distribution peaked around the expected value $\overline{s^*}=1$. However, in order to investigate whether this distribution is `narrow enough' in order to achieve the desired reconstructability of the network, we have to look more closely at the relative fluctuations $\delta_i(z_\mathrm{sparse})$ for each node $i$.
\begin{figure}[t!]
     \centering
     \begin{subfigure}[b]{0.45\textwidth}
         \centering
         \includegraphics[width=\textwidth]{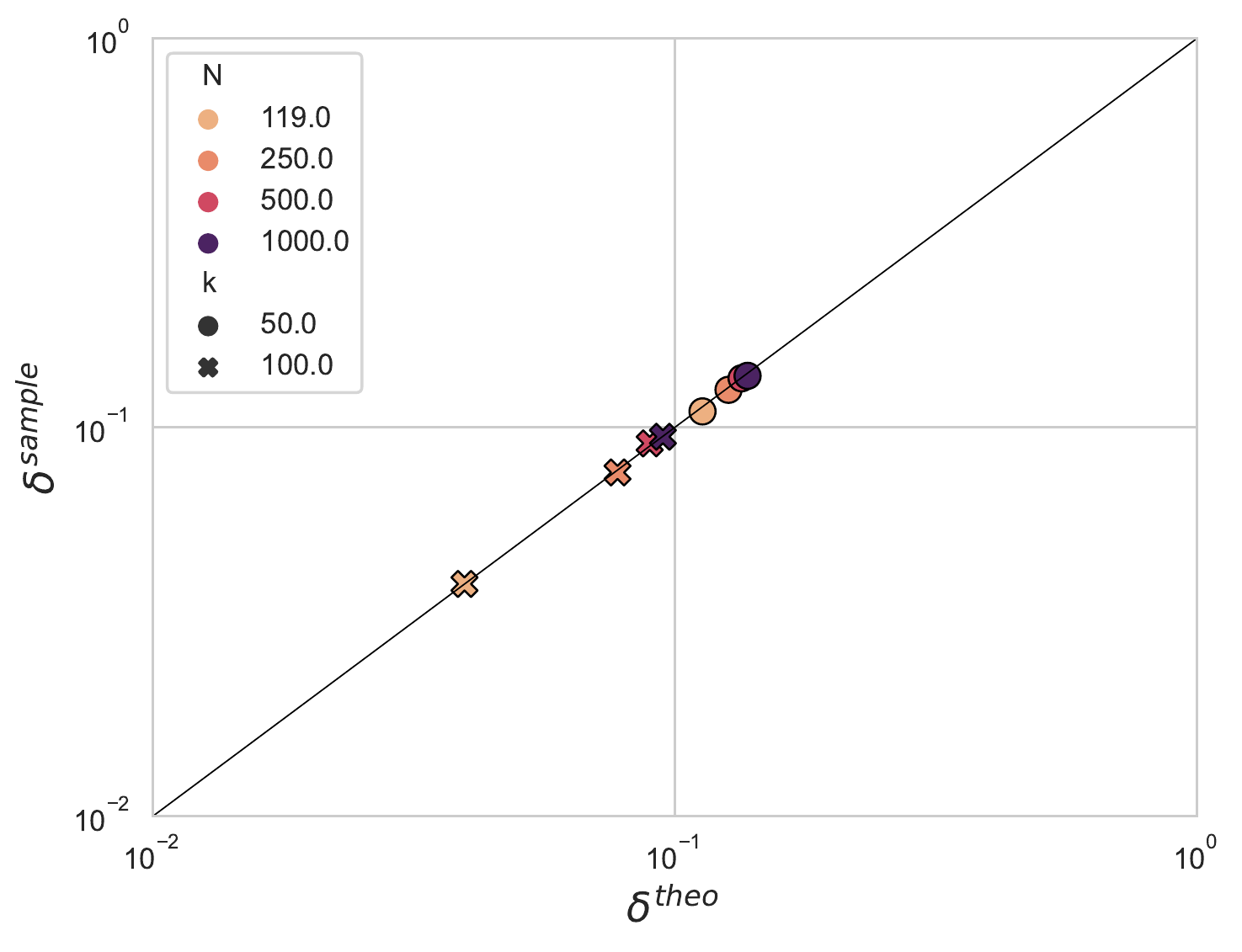}
         \caption{\label{equal_deltatheogen}}
     \end{subfigure}
     \hfill
     \begin{subfigure}[b]{0.45\textwidth}
         \centering
         \includegraphics[width=\textwidth]{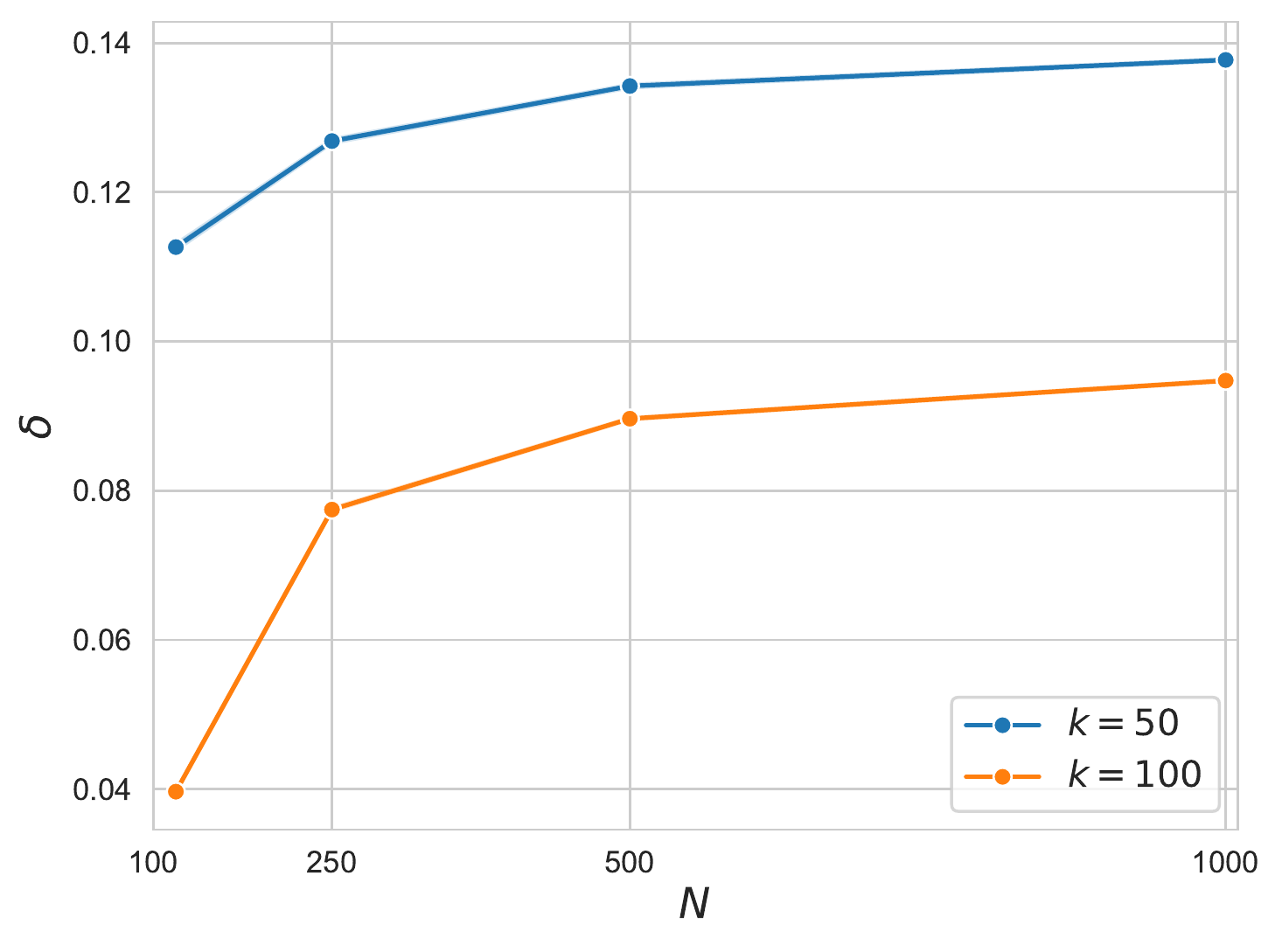}
         \caption{\label{equal_deltaN}}
     \end{subfigure}
        \caption{Relative strength fluctuations $\delta_i(z_\mathrm{sparse})$ in the case of identical nodes (homogeneous networks), for a randomly chosen node $i$.
        (a) The theoretical value $\delta_i^\mathrm{theo}(z_\mathrm{sparse})$ given by Eq.~\eqref{relative} matches the realized value $\delta_i^\mathrm{sample}(z_\mathrm{sparse})$ calculated across 1000 sampled networks, for different values of $N$ and $k$.
        (b)
        Realized $\delta_i^\mathrm{sample}(z_\mathrm{sparse})$ versus the number $N$ of nodes, which asymptotically approaches the theoretical value $1/\sqrt{k}$.
        }
        \label{fig:equal_delta}
\end{figure}

This is what we illustrate in Fig.~\ref{equal_deltatheogen}, where for one randomly chosen node (note that all nodes are statistically equivalent in the homogeneous case) we compare the theoretical value $\delta_i^\mathrm{theo}(z_\mathrm{sparse})$ calculated in Eq.~\eqref{relative} with the sample fluctuations $\delta_i^\mathrm{sample}(z_\mathrm{sparse})$ measured using the sample variance across the 1000 realizations of the network, for different values of $N$ and $k$. 
We confirm that sample and theoretical values are in excellent agreement. In Fig.~\ref{equal_deltaN} we also confirm that, after a transient trend that increases with $N$, the relative fluctuations approach the asymptotically constant value $\delta_i(z_\mathrm{sparse})={1}/{\sqrt{k}}$ for all $i$, as expected from Eq.~\eqref{1/k}, and hence they do not vanish. 
Indeed, as discussed in Sec.~\ref{homo}, the relative fluctuations for homogeneous networks decay only in the dense regime.

\begin{figure}[b!]
\centering
\includegraphics[width=1\textwidth]{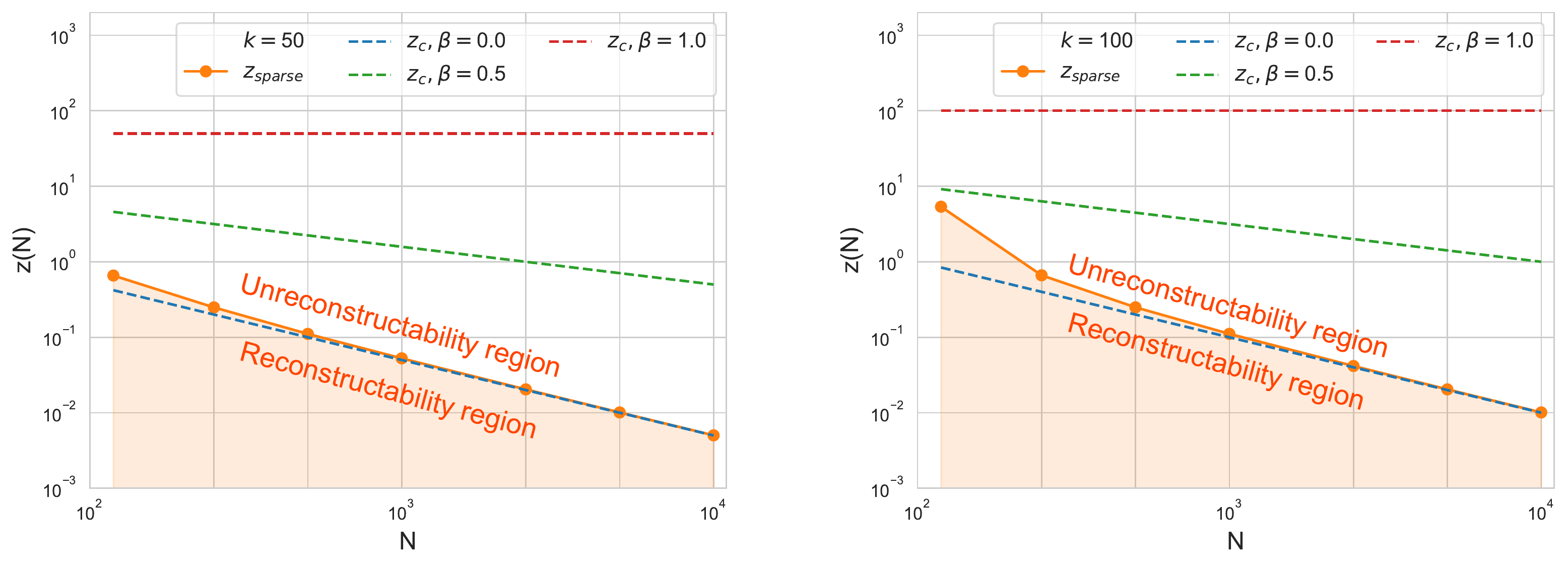}
\caption{Comparison between $z_\mathrm{sparse}(N)$ and $z_c(N)\simeq{c}^{-1} N^{\beta -1}$ with $c=1/k$, for different values of $\beta$ and $N$ in the case of homogeneous networks. Each subplot corresponds to a different choice of the constant $k$. The shaded area is the region $z(N)<z_\mathrm{sparse}(N)$: the network is reconstructable whenever $z_c(N)$ crosses that region. This occurs only for $\beta\to 0^+$ and small $N$, while for large $N$ the network is always in the unreconstructability regime.}
\label{fig:equal_zeti}
\end{figure}

In Fig.~\ref{fig:equal_zeti} we compare the behaviour of $z_\mathrm{sparse}(N)$, obtained by inverting Eq.~\eqref{sparse} (for $k=50$ and $k=100$), with the critical $z_c(N)$ given by Eq.~\eqref{zc_ER}, with the identification $c=1/(k\,\mathrm{min}_i\{s_i^*\})=1/k$ that would correspond to the baseline asymptotic behaviour for $\delta(z_\mathrm{sparse})$ when $\beta\to 0^+$.  The figure confirms that, asymptotically, the network is unreconstructable ($z_c>z_\mathrm{sparse}$) for all the values $0<\beta<1$ allowed by Eq.~\eqref{zc_ER}.
Only for small $\beta\to 0^+$ and moderate values of $N$ the network can be found transiently in the reconstructability phase. 

\subsubsection{Bankscope strengths.}
We then consider the heterogeneous case where the strengths of banks are either taken from the Bankscope data or sampled from the fitted distribution with CDF given by Eq.~\eqref{CDF}, as discussed in Sec.~\ref{data_description}. We enforce the same values of density as in the previous example (sparse regime) and vary $N$ and $k$ as before.

\begin{figure}[t!]
\centering
\includegraphics[width=1\textwidth]{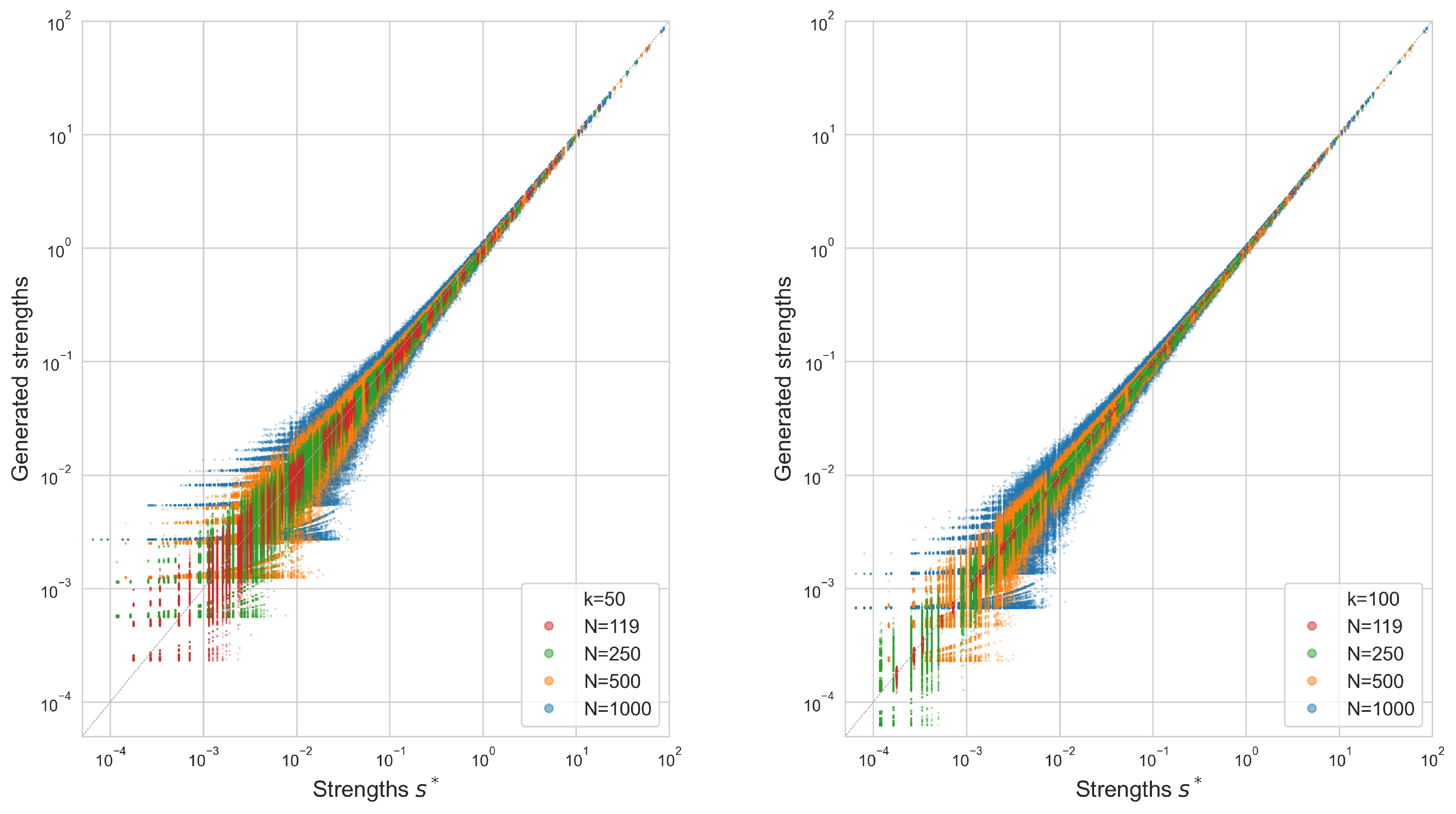}
\caption{Empirical strengths $s^*$ versus the generated ones. Each dot figures out the value of the generated strength of a certain node and a certain network between the 1000 realizations in the ensemble versus its empirical strength. Each subplot corresponds to a different choice of the constant $k$ and different colors correspond to a different number of nodes $N$. }
\label{fig:realvsindividual}
\centering
\end{figure}

Figure~\ref{fig:realvsindividual} reports a scatter plot of the realized node strength of each node $i$ (in a typical realization out of the 1000 sampled ones) versus the corresponding empirical strength $s^*_i$, which is also the expected value across the entire ensemble. We see that, in line with our theoretical calculations, nodes with smaller strength feature larger relative displacements from the expected identity line. For a fixed value of $k$, the relative displacement increases as the number $N$ of nodes increases (hence as the link density decreases). Larger displacements correspond to a worse reconstruction. 

\begin{figure}[t!]
\centering
\includegraphics[width=1\textwidth]{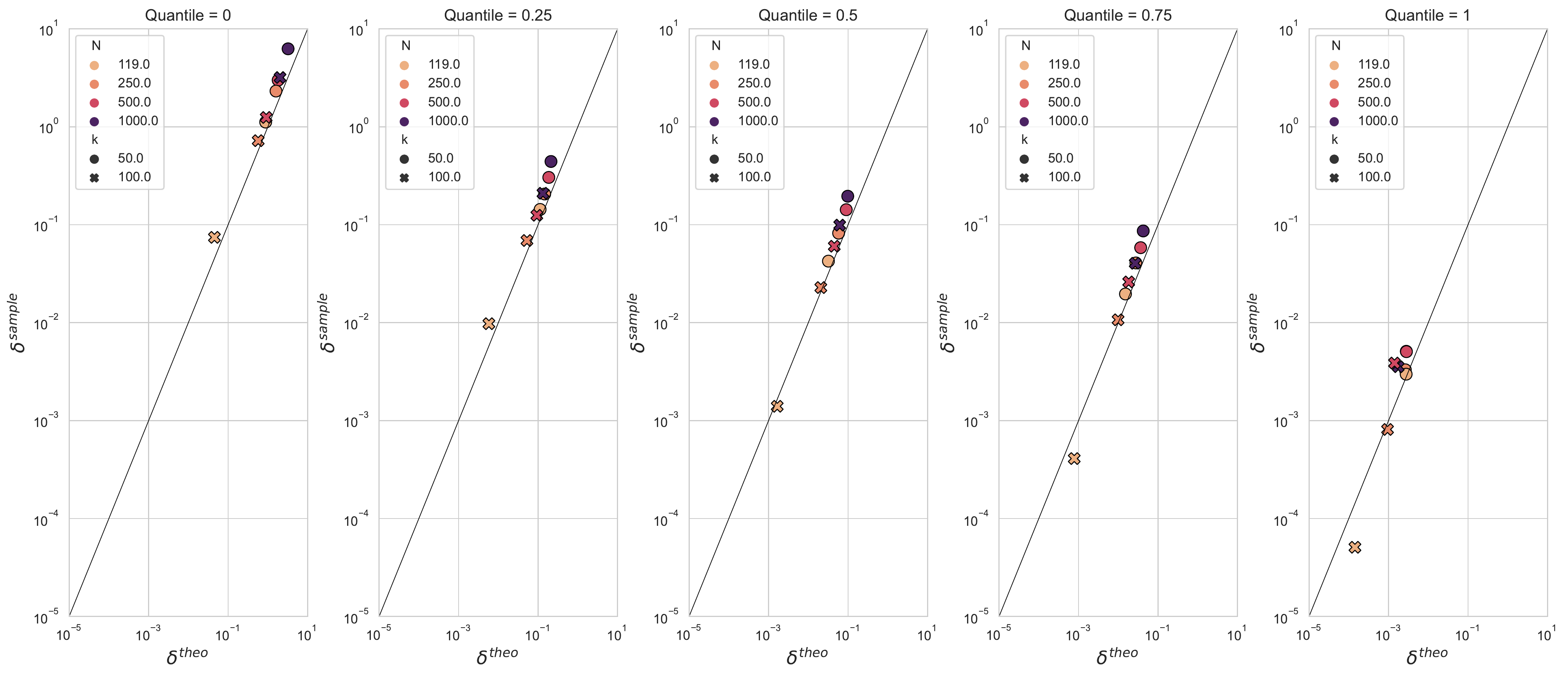}
\caption{
Relative strength fluctuations $\delta_i(z_\mathrm{sparse})$ in the reconstructed networks based on the Bankscope data (heterogeneous networks).
The theoretical value $\delta_i^\mathrm{theo}(z_\mathrm{sparse})$ given by Eq.~\eqref{relative} is compared with the realized value $\delta_i^\mathrm{sample}(z_\mathrm{sparse})$ calculated across 1000 sampled networks, for different values of $N$ and $k$. Each subplot shows the fluctuations for a representative node that corresponds to the indicated quantile of the strength distribution.}
\label{fig:deltaobsandtheo}
\end{figure}

To quantify the above effect rigorously, in Fig.~\ref{fig:deltaobsandtheo} we focus on the relative fluctuations $\delta_i(z_\mathrm{sparse})$ and, in analogy with Fig.~\ref{equal_deltatheogen}, compare the sample fluctuations $\delta_i^\mathrm{sample}(z_\mathrm{sparse})$ (measured using the sample variation across 1000 realizations of the network) with the theoretical value $\delta_i^\mathrm{theo}(z_\mathrm{sparse})$ calculated in Eq.~\eqref{relative}, for different values of $N$ and $k$. 
Since in this case the banks' strengths vary greatly in size, we report in different subplots the fluctuations for different representative nodes, each corresponding to a certain quantile of the strength distribution. 
Again, we confirm the good accordance between the sampled fluctuations and our theoretical calculations, for all quantiles.

\begin{figure}[t!]
\centering
\includegraphics[width=1\textwidth]{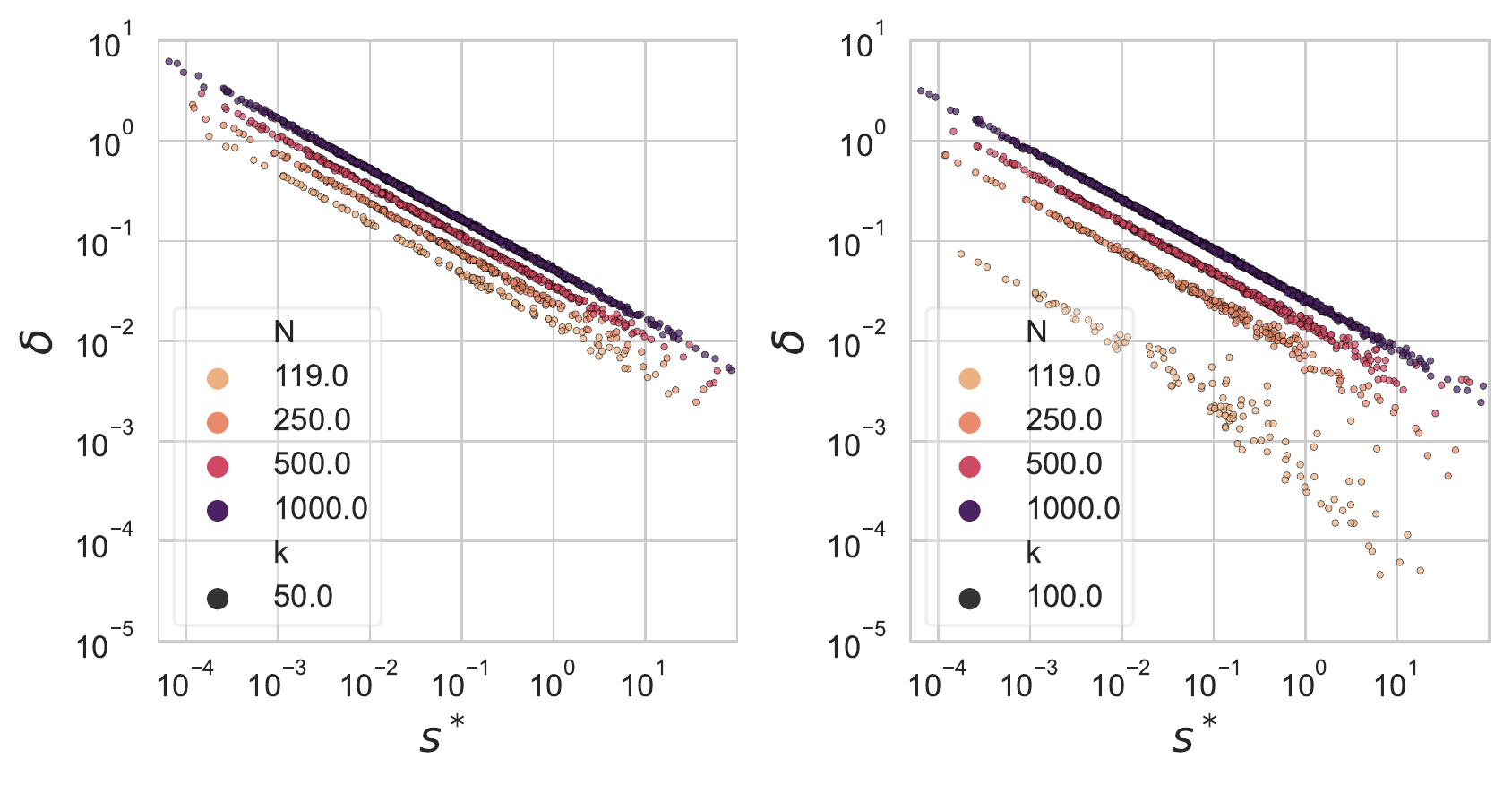}
\caption{Sample relative  fluctuation $\delta_i^\mathrm{sample}(z_\mathrm{sparse})$ versus the empirical strength $s^*_i$.
Each subplot corresponds to a different value of $k$, and different colors correspond to a different number $N$ of nodes.}
\label{fig:strengthvsfluctuations}
\centering
\end{figure}

\begin{figure}[b!]
\centering
\includegraphics[width=0.5\textwidth]{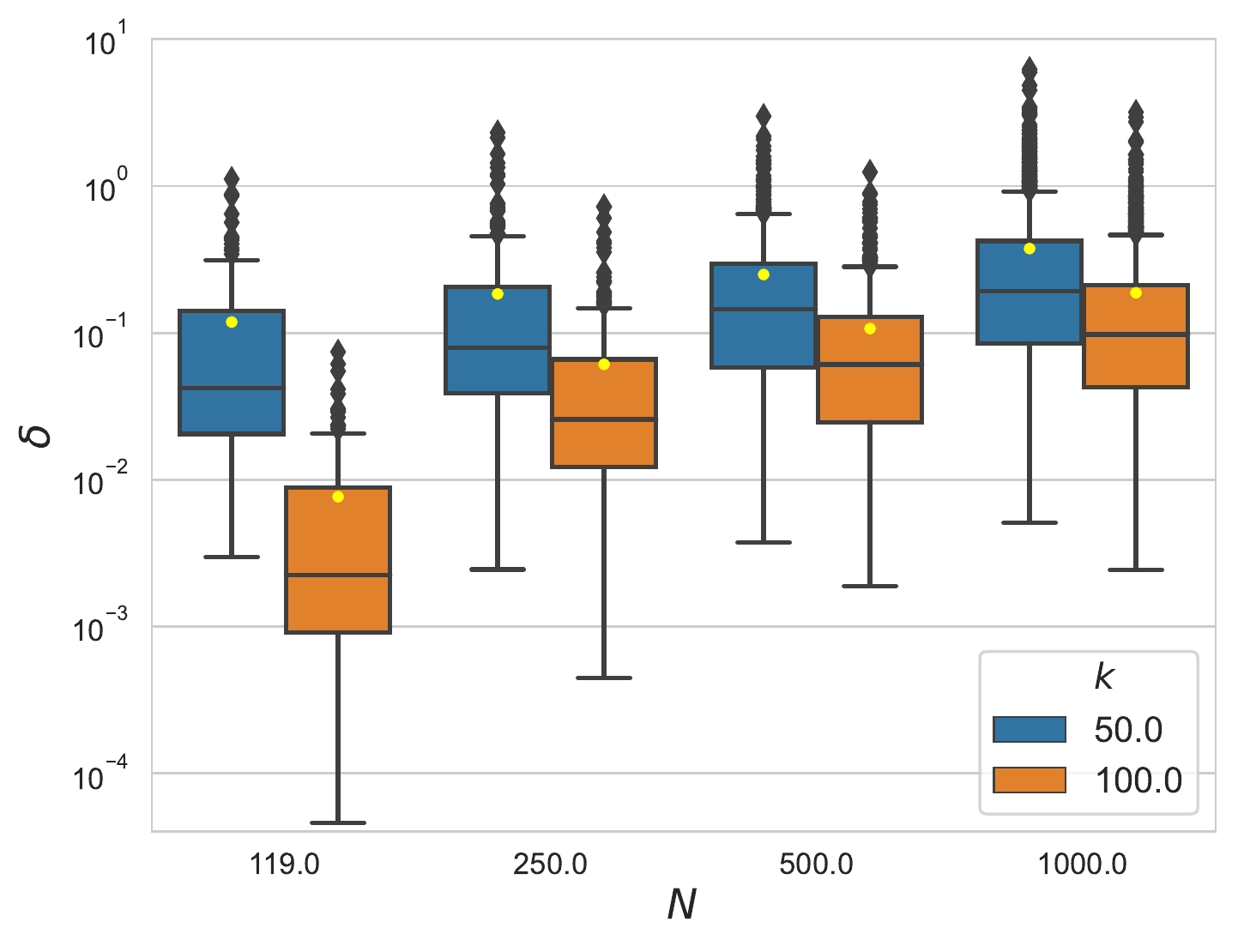}
\caption{Boxplot representing the distribution of the sample relative fluctuations $\delta_i^\mathrm{sample}(z_\mathrm{sparse})$ over nodes, as a function of the number $N$ of nodes. Yellow dots represent the mean. Different colors correspond to different values of $k$.}
\label{fig:Nvsdelta}
\centering
\end{figure}

As a related check, Fig.~\ref{fig:strengthvsfluctuations} shows the sample relative fluctuation $\delta_i^\mathrm{sample}(z_\mathrm{sparse})$ versus the empirical strength $s^*_i$.
For fixed $k$ and $N$, $\delta_i^\mathrm{sample}(z_\mathrm{sparse})$ is proportional to $1/\sqrt{s^*_i}$, in accordance with the expectation from Eq.~\eqref{relative}.
Moreover we see that, for fixed $s^*_i$, the dependence of $\delta_i^\mathrm{sample}(z_\mathrm{sparse})$ on $N$ is not suppressed. This is different from the behaviour we found in Eq.~\eqref{sparse_ER} for the homogeneous case, where the dependence of $\delta_i^\mathrm{sample}(z_\mathrm{sparse})$ on $N$ is exactly cancelled out by the expression $z_\mathrm{sparse}\simeq k/N$ given by Eq.~\eqref{1/k}, which in this case evidently does not hold (at least for the considered range of values for $N$). In particular, here we see that $\delta_i^\mathrm{sample}(z_\mathrm{sparse})$ increases with $N$, a behaviour that is confirmed in Fig.~\ref{fig:Nvsdelta}. 

\begin{figure}[b!]
\centering
\includegraphics[width=1\textwidth]{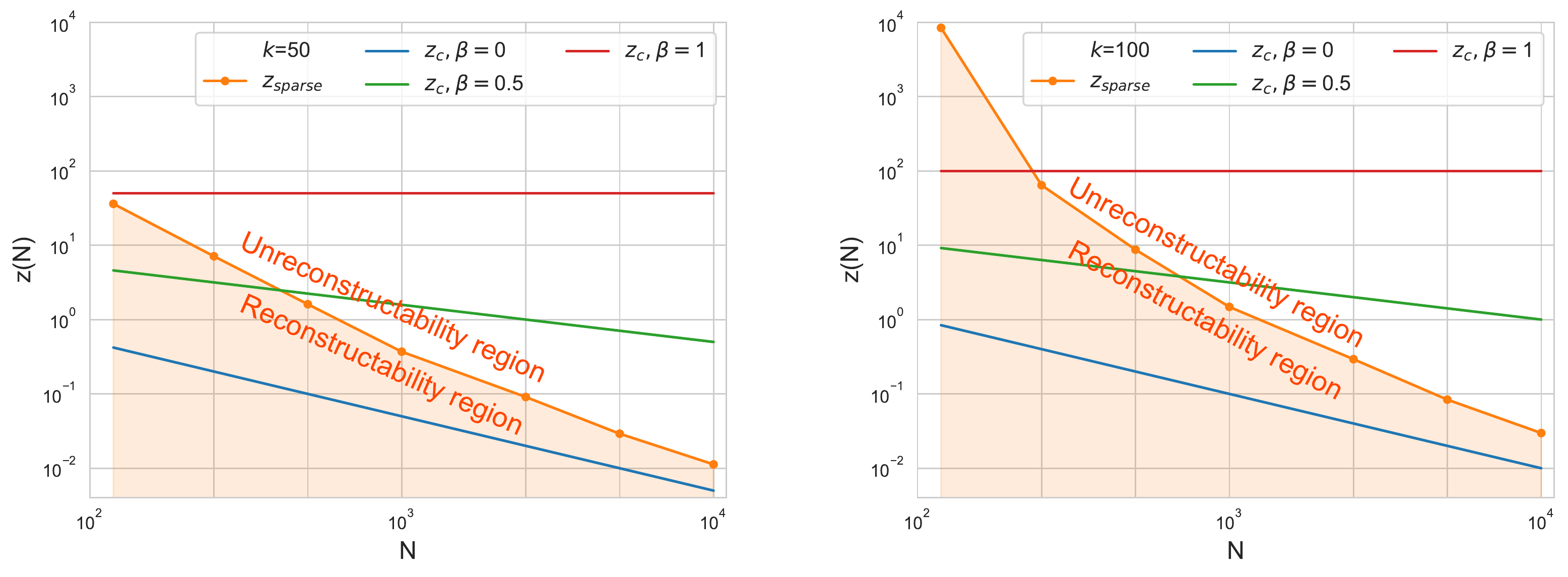}
\caption{
Comparison between $z_\mathrm{sparse}(N)$ and $z_c(N)\simeq{c}^{-1} N^{\beta -1}$ with $c=1/(k\,\min_i\{s_i^*\})$, for different values of $\beta$ and $N$ in the case of networks with strengths based on the Bankscope data. Each subplot corresponds to a different choice of the constant $k$. The shaded area is the region $z(N)<z_\mathrm{sparse}(N)$, and the network is reconstructable whenever $z_c(N)$ crosses that region. Whether this occurs depends strongly on $\beta$: for $\beta\to 0^+$ the network is reconstructable for all values of $N$, for $\beta=1$ the network is always unreconstructable, while for intermediate values of $\beta$ the reconstructability depends on $N$.}
\label{fig:bankscope_zeti}
\centering
\end{figure}

In Fig.~\ref{fig:bankscope_zeti} we compare $z_\mathrm{sparse}(N)$ and $z_c(N)\simeq{c}^{-1} N^{\beta -1}$ with $c^{-1}=k\,\min_i\{s_i^*\}$, which again would correspond to the baseline decay of the relative fluctuations, for different values of $\beta$ and $N$.
We find the presence of two regimes (reconstructability and unreconstructability), depending on the combination of values of $\beta$ and $N$. So, in this case the reconstructability depends sensibly on the details of the strength distribution and on the size of the network. Even if certain values of $\beta$ lead asymptotically to unreconstructability, we find that for finite but realistically large $N$ the system may still be in the reconstructability regime. 
In other words,  depending on the chosen values for $k$ and $\beta$, the networks are reconstructable up to a critical number of nodes whose value increases as $\beta$ decreases (hence as the decay of the largest relative fluctuations of the strength slows down). 

\subsubsection{Self-loops contribution.}
We finally confirm that, as expected, the effect of the induced self-loops in the reconstruction is negligible. This is shown in Table \ref{table:wloops} where we report the fraction of weight associated to self-loops, confirming that the condition in Eq.~\eqref{eq:condition} is met, as the fraction decreases for increasing $N$. 

\begin{table}[h!]
\centering
\begin{tabular}{|c|c|c|}
\toprule
 N  & Homogeneous networks & Heterogeneous networks  \\
 \midrule
 119 &  0.008 &  0.061 \\
250 &  0.004 & 0.073 \\
 500 &  0.002 & 0.049 \\
1000 &  0.001 & 0.026 \\
\bottomrule  
\end{tabular}
\vskip 5pt
\caption{Fraction ${\langle W_{SL}\rangle}/{\langle W\rangle}$ of total link weight associated to the self-loops induced by the reconstruction method, for different numbers of nodes $N$ in the case of homogeneous (equal values) and heterogeneous (from Bankscope data) strength distributions.}
\label{table:wloops}
\end{table}

\section{Conclusions\label{sec:conclu}}
In this paper we focused on the reconstruction of financial networks from aggregate constraints (node strengths, representing assets and liabilities) and introduced the concept of \textit{reconstructability}, occurring when the constraints, besides being reproduced on average, are also close to their expected value in individual typical realizations of the ensemble. 
We considered different situations arising in the sparse regime, first from a theoretical point of view and then by generating networks from real-world strength distributions. 
In the homogeneous case of equal strengths, we found that simultaneously sparse and large networks are always unreconstructable.
By contrast, if an appropriate degree of heterogeneity is introduced (specifically, a core-periphery structure), we found that the system can be in one of two regimes (reconstructability and unreconstructability), depending on the asymptotic decay of the minimum strength.
In general, the behaviour of the minimum strength plays a crucial role in the reconstruction.
Using data from the Bankscope dataset, and extrapolating to a larger number of nodes,  we found again the presence of two regimes, which additionally depend sensibly on the number of nodes and on the details of the strength distribution.

It should be noted that the independence of different links captured by the reconstruction method discussed in Sec.~\ref{reconstruction}, together with the fact that both the connection probability $p_{ij}$ in Eq.~\eqref{p} and the link weight $w_{ij}$ in Eq.~\eqref{w} are determined by purely local properties ($s^*_i$ and $s^*_j$), indicates that it is possible to use the reconstruction method as a form of a decentralized market clearing mechanism where all the constraints are met simultaneously via purely pairwise interbank interactions, as long as the reconstructability criterion in Eq.~\eqref{condition} is met. Violations of the reconstructability condition make market clearing more difficult without a centralized entity capable of enforcing all constraints globally. Therefore the existence of a critical density for reconstructability implies that central bank interventions that lower the density of links below the critical threshold may unintentionally favour a liquidity crisis.
More in general, if individual realizations of the reconstructed networks are used by regulators as substrates to simulate the propagation of shocks throughout the interbank system, a mismatch between the realized marginals and the empirical ones could lead to an incorrect estimation of systemic risk.
Reconstructability then becomes an important criterion to be met in order to avoid the resulting bias.

Our result suggest that network reconstructability is an important aspect of probabilistic reconstruction techniques and deserves further study, including its generalization to directed and more complicated network structures.

\subsubsection{Acknowledgements} 
DG is supported by the Dutch Econophysics Foundation (Stichting Econophysics, Leiden). This work is supported by the European Union - NextGenerationEU - National Recovery and Resilience Plan (Piano Nazionale di Ripresa e Resilienza, PNRR), project `SoBigData.it - Strengthening the Italian RI for Social Mining and Big Data Analytics' - Grant IR0000013 (n. 3264, 28/12/2021). We also acknowledge the PRO3 project ``Network Analysis of Economic and Financial Resilience'' by the IMT School of Advanced Studies Lucca, the Scuola Normale Superiore in Pisa and the Sant'Anna School of Advanced Studies in Pisa.
\bibliographystyle{unsrt}
\bibliography{bibliography}

\end{document}